\documentclass[twocolumn,superscriptaddress,floatfix,prc,amsmath,amssymb,nofootinbib]{revtex4}
\usepackage{graphicx}
\usepackage{dcolumn}
\usepackage{bm}
\usepackage{gensymb}
\usepackage{siunitx}
\usepackage{tikz}

\usepackage{color,hyperref}
\hypersetup{colorlinks,breaklinks,linkcolor=blue,urlcolor=blue,anchorcolor=blue,citecolor=blue}

\usepackage[english]{babel}

\usepackage{multirow}
\usepackage{orcidlink}
\usepackage{color}
\usepackage{soul}
\usepackage{rotating}
\usepackage{longtable}
\usepackage{footnote}
\usepackage{setspace}
\usepackage{afterpage}
\begin{document}

\title{High-spin spectroscopy and the onset of quasicollective structures in $^{69}$Ga}

\author{F. E. Idoko\,\orcidlink{0000-0003-2024-9535}}
 \email{fidoko@unc.edu}
 \affiliation{Department of Physics and Astronomy, University of North Carolina Chapel Hill, NC 27599, USA}
\affiliation{Triangle Universities Nuclear Laboratory, Duke University, Durham, NC 27708, USA}

\author{A. D. Ayangeakaa\,\orcidlink{0000-0003-1679-3175}}
 \affiliation{Department of Physics and Astronomy, University of North Carolina Chapel Hill, NC 27599, USA}
\affiliation{Triangle Universities Nuclear Laboratory, Duke University, Durham, NC 27708, USA}

\author{N. Sensharma\,\orcidlink{0000-0002-5046-9451}}
\altaffiliation[Present Address: ]{Physics Division, Argonne National Laboratory, Lemont, Illinois 60439, USA}
 \affiliation{Department of Physics and Astronomy, University of North Carolina Chapel Hill, NC 27599, USA}
\affiliation{Triangle Universities Nuclear Laboratory, Duke University, Durham, NC 27708, USA}

\author{C. J. Chiara\,\orcidlink{0000-0001-9969-0248}}
\affiliation{U.S. Army Research Laboratory, Adelphi, Maryland 20783, USA.}

\author{S. Zhu}\thanks{Deceased}
\affiliation{National Nuclear Data Center, Brookhaven National Laboratory, Upton, New York 11973-5000, USA.}

\author{E. A. McCutchan}
\affiliation{National Nuclear Data Center, Brookhaven National Laboratory, Upton, New York 11973-5000, USA.}

\author{A. Saracino\,\orcidlink{0009-0002-0357-2429}}
 \affiliation{Department of Physics and Astronomy, University of North Carolina Chapel Hill, NC 27599, USA}
\affiliation{Triangle Universities Nuclear Laboratory, Duke University, Durham, NC 27708, USA}

\author{R. V. F. Janssens\,\orcidlink{0000-0001-7095-1715}}
 \affiliation{Department of Physics and Astronomy, University of North Carolina Chapel Hill, NC 27599, USA}
\affiliation{Triangle Universities Nuclear Laboratory, Duke University, Durham, NC 27708, USA}


\author{H. M. Albers}
\affiliation{GSI Helmholtzzentrum für Schwerionenforschung GmbH, 64291 Darmstadt, Germany.}

\author{S. Balderrama}
\affiliation{National Nuclear Data Center, Brookhaven National Laboratory, Upton, New York 11973-5000, USA.}

\author{L. Canete}
\affiliation{School of Mathematics and Physics, University of Surrey, Guildford GU2 7XH, United Kingdom}

\author{J. Carroll}
\affiliation{U.S. Army Research Laboratory, Adelphi, Maryland 20783, USA.}

\author{M. P. Carpenter}
\affiliation{Physics Division, Argonne National Laboratory, Lemont, Illinois 60439, USA}

\author{P. A. Copp\,\orcidlink{0000-0002-4786-2404}}
\affiliation{Los Alamos National Laboratory: Los Alamos, New Mexico, USA}

\author{D. T. Doherty}
\affiliation{School of Mathematics and Physics, University of Surrey, Guildford GU2 7XH, United Kingdom}

\author{P. Golubev}
\affiliation{Department of Physics, Lund University, 22100 Lund, Sweden}

\author{D. J. Hartley}
\affiliation{Department of Physics, U.S. Naval Academy, Annapolis, Maryland 21402, USA}

\author{A. B. Hayes}
\affiliation{National Nuclear Data Center, Brookhaven National Laboratory, Upton, New York 11973-5000, USA.}

\author{Y. Hrabar\,\orcidlink{0000-0002-3194-1001}}
\affiliation{Department of Physics, Lund University, 22100 Lund, Sweden}


\author{H. Jayatissa}
\affiliation{Physics Division, Argonne National Laboratory, Lemont, Illinois 60439, USA}

\author{M. Miranda}
\affiliation{National Nuclear Data Center, Brookhaven National Laboratory, Upton, New York 11973-5000, USA.}

\author{C. Müller-Gatermann\,\orcidlink{0000-0002-9181-5568}}
\affiliation{Physics Division, Argonne National Laboratory, Lemont, Illinois 60439, USA}

\author{N. N. O'Briant}
 \affiliation{Department of Physics and Astronomy, University of North Carolina Chapel Hill, NC 27599}

\author{M. Siciliano}
\affiliation{Physics Division, Argonne National Laboratory, Lemont, Illinois 60439, USA}

\author{F. G. Kondev}
\affiliation{Physics Division, Argonne National Laboratory, Lemont, Illinois 60439, USA}

\author{T. M. Kowalewski\,\orcidlink{0000-0001-7493-1842}}
 \affiliation{Department of Physics and Astronomy, University of North Carolina Chapel Hill, NC 27599}
\affiliation{Triangle Universities Nuclear Laboratory, Duke University, Durham, NC 27708, USA}

\author{T. Lauritsen}
\affiliation{Physics Division, Argonne National Laboratory, Lemont, Illinois 60439, USA}

\author{W. Reviol}
\affiliation{Physics Division, Argonne National Laboratory, Lemont, Illinois 60439, USA}

\author{D. Rudolph\,\orcidlink{0000-0003-1199-3055}}
\affiliation{Department of Physics, Lund University, 22100 Lund, Sweden}

\author{J. Rufino}
\affiliation{National Nuclear Data Center, Brookhaven National Laboratory, Upton, New York 11973-5000, USA.}

\author{D. Seweryniak}
\affiliation{Physics Division, Argonne National Laboratory, Lemont, Illinois 60439, USA}

\author{J. R. Vanhoy}
\affiliation{Department of Physics, U.S. Naval Academy, Annapolis, Maryland 21402, USA}

\author{W. B. Walters\,\orcidlink{0000-0001-5692-6943}}
    \affiliation{
        Department of Chemistry and Biochemistry, University of Maryland College Park, College Park, Maryland 20742, USA
        }

\author{G. L. Wilson}
\affiliation{Physics Division, Argonne National Laboratory, Lemont, Illinois 60439, USA}
\affiliation{Louisiana State University, Baton Rouge, Louisiana 70803, USA}


\date{\today}

\begin{abstract}

\noindent \textbf{Background:}  
Nuclei in the mass $A \approx 60$--70 region exhibit a rich interplay between single-particle and collective excitations, particularly near the $N=40$ subshell closure where deformation-driving orbitals such as $g_{9/2}$ begin to influence the structure. In odd-mass gallium isotopes, the evolution from spherical to deformed configurations with increasing spin remains an active area of study.

\noindent \textbf{Purpose:}  
This study aims to investigate the evolution of nuclear structure in the odd-mass nucleus $^{69}$Ga, with emphasis on the emergence of collective excitations at high spin. Although nuclei in the vicinity of $N=40$ exhibit deformation driven by $g_{9/2}$ orbital occupancy, the degree to which such behavior develops in $^{69}$Ga has not been well established. By extending the level scheme and identifying new high-spin structures, the present work seeks to clarify the role of high-$j$ orbitals and provide benchmarks for theoretical models describing collectivity in this mass region.

\noindent \textbf{Method:}  
High-spin states in $^{69}$Ga were populated using the fusion-evaporation reaction $^{26}$Mg($^{48}$Ca,\,$p4n\gamma$) at a beam energy of 195 MeV. The experiment was carried out at the Argonne National Laboratory using the Gammasphere multidetector array in conjunction with the Fragment Mass Analyzer. Coincidence relationships between the emitted $\gamma$ rays were used to construct the level scheme, while spin assignments were established through angular distribution and correlation measurements. Shell-model and tilted-axis cranking covariant density functional theory (TAC-CDFT) calculations were employed to interpret the observed structures.

\noindent \textbf{Results:}  
The level scheme of $^{69}$Ga was significantly extended up to an excitation energy of $\sim 15$ MeV and spin $45/2\hbar$. At low spins, the structure is well described by single-particle excitations and is satisfactorily reproduced by shell-model calculations using the JUN45 and jj44b interactions. At higher spins ($J \geq 21/2$~$\hbar$), three sequences of stretched $E2$ transitions were identified, indicating the onset of quasicollective behavior. The TAC-CDFT calculations suggest that these bands originate from aligned configurations involving both proton and neutron $g_{9/2}$ orbitals, thus highlighting the emergence of deformation in $^{69}$Ga at high angular momentum.

\noindent \textbf{Conclusions:}
This work extends the known level structure of $^{69}$Ga and provides evidence for the development of collective behavior at high spin. While low-spin states are well described by shell-model calculations and dominated by single-particle excitations, the emergence of stretched $E2$ sequences above $21/2\hbar$ indicates the onset of quasicollective rotational motion. These results reveal the influence of high$j$ orbitals near $N=40$ and provide valuable input to understand the structure evolution of odd-mass nuclei in the $A \approx 70$ region.

\end{abstract}

\maketitle

\section{INTRODUCTION}  
The spectral behavior of nuclei in the $A \approx 60-70$ mass region, on both sides of the valley of beta stability, has been the focus of extensive experimental and theoretical investigations for many years. These nuclei, with a relatively small number of particles in the valence space, exhibit a variety of nuclear phenomena and, hereby, provide an ideal laboratory for exploring the interplay between single-particle and collective dynamics. Generally, near the ground state, their structure is characterized predominantly by single-particle excitations involving the $1f_{5/2}$, $2p_{3/2}$, and $2p_{1/2}$ spherical shell-model states. However, at higher angular momenta and excitation energies, long-range correlations between valence nucleons result in the emergence of deformed nuclear shapes and collective behavior. In addition, multiparticle-multihole excitations from the $1f_{7/2}$ orbital (below the $N,Z = 28$ spherical shell gaps) to the $1g_{9/2}$ high-$j$ orbital (above these gaps), contribute to the appearance of deformed minima in the potential energy surface. This complex interplay results in a range of phenomena, such as shape coexistence~\cite{PhysRevLett.82.3763,PhysRevC.85.064305} and rapid shape changes~\cite{Andreoiu59Cu2002}. Furthermore, configuration-dependent pairing correlations and highly-deformed intruder configurations of rotational sequences, involving proton and neutron particle-hole excitations, in some instances $1p-1h$, $1p-2h$, $2p-2h$ and higher-order configurations have been observed \cite{ward2000,rudolph1998,yu2002,johansson2009,Rudolph_2010}. 
Within this mass region, these structural effects are highly sensitive to small changes in nucleon number, emphasizing the importance of detailed investigations across select isotopic/isotonic chains.

Over the years, systematic studies focusing on the evolution of structure across the region have been undertaken, most extensively in the even-even nuclei \cite{Cr,64cr,PhysRevC.85.044316-Steppenbeck,59-60Fe,60-62Fe,64fe,62Ni,albers63Ni,PhysRevC.91.044327, PhysRevC.105.044315,66zn,62Cu}. However, the odd-mass ones present additional layers of complexity due to the significant impact of the unpaired nucleon. Studies of these odd systems have been valuable in delineating the evolution of single-particle levels around the $Z=28$ closed shell and in exploring the coupling of the quasiparticle degrees of freedom to the low-frequency quadrupole vibrational field of the even-even core through multipole-multipole forces \cite{Bakoyeorgos1982}. In many of these nuclei, this coupling has been systematically associated with the coexistence and mixing of quasivibrational and quasirotational structures at low and medium spins, as demonstrated in the case of the $Z=31$ odd-even Ga isotopes. 
With a three valence proton cluster outside the $Z=28$ closed shell, these isotopes present a relatively complex structure due to the inclusion of a broken proton pair in collective effects. 

Low-lying states in odd-mass Ga nuclei have been investigated using a variety of experimental techniques including proton transfer, Coulomb excitation and other scattering experiments as well as radioactive decay \cite{Paradellis1981,Harms-Ringdahl1974,IVASCU1974,Diriken2010, Stefanescu2009}. The observed level sequences have been interpreted in terms of single-particle and core-coupled configurations involving the $f_{7/2}$, $p_{3/2}$, $f_{5/2}$ and $p_{1/2}$ proton orbitals. For example, results from $(^{3}\mathrm{He},d)$ reactions on $^{64,66,68,70}\mathrm{Zn}$ targets showed significant shifts in transition strengths and excitation energies for low-lying states in $^{71}\mathrm{Ga}$ when compared to $^{65,67,69}\mathrm{Ga}$ \cite{RICCATO1974}. Specifically, the $\ell=1$ transition strength to the first excited state in $^{71}\mathrm{Ga}$ was found to be an order of magnitude smaller than in the lighter Ga isotopes. In addition, the excitation energies of the $\ell=3$ ($5/2^-$) and $\ell=4$ ($9/2^+$) levels decreased in $^{71}\mathrm{Ga}$. The observed discontinuities at $N=40$ were attributed to an increased proton-neutron interaction between particles occupying the same orbital. This deviation was confirmed by a $(d,^{3}\mathrm{He})$ transfer reaction, which indicated abrupt changes in proton configurations between $N=40$ and 42. The sharp rise in excitation energy of the $1/2^-$ state from $320$ $\mathrm{keV}$ in $^{69}\mathrm{Ga}$ to 1109 keV in $^{71}\mathrm{Ga}$ exemplifies this shift \cite{RICCATO1974, rotbard1978}.

On the other hand, high-spin levels in neutron-deficient Ga nuclei have been obtained from various heavy-ion induced reactions and shown to exhibit characteristics of particle-core coupling, suggesting moderate prolate deformation consistent with theoretical predictions. Specifically, the $^{40}$Ca($^{32}$S, $2\alpha\gamma$) and $^{40}$Ca($^{32}$S $\alpha3p\gamma$) fusion evaporation reactions, performed using the GASP $\gamma$-ray spectrometer in conjunction with the ISIS charged-particle detector system, were used to study level structures in $^{63,65}$Ga \cite{Weiszflog2001}. The reported low-lying negative-parity states of single-particle character were attributed to $(\pi{p}_{3/2} \otimes 2^+)$ and $(\pi{f}_{5/2} \otimes 2^+)$ couplings yielding ${7/2}^-$ and ${9/2}^-$ states. Moreover, a rotational-like sequence with positive parity was observed atop the ${9/2}^+$ level in both nuclei and was interpreted as originating from the alignment of a pair of $g_{9/2}$ neutrons. 
Similarly, the $^{12}$C($^{58}$Ni, $\alpha p$) and $^{12}$C($^{58}$Ni, $3p$) fusion-evaporation reactions were employed in Ref.~\cite{danko1999} to investigate the nature of positive- and negative-parity states in $^{65,67}$Ga and these were interpreted within the framework of the interacting boson-fermion model. Notably, the negative-parity levels observed below $\approx 4$ MeV were associated with the coupling of a quasiproton to quadrupole-phonon states, while the positive-parity levels were attributed specifically to the coupling of proton particles in the $g_{9/2}$ orbital with quadrupole phonon states. Further investigations of the structure of heavier odd-mass Ga isotopes around $A \approx 70$ have also been performed. Stefanescu \textit{et al.}~\cite{Stefanescu2009} populated excited states in $^{71,73,75,77}$Ga via deep-inelastic reactions of a $^{76}$Ge beam on a $^{238}$U target, reporting energy levels built upon the ${3/2}^-$, ${7/2}^-$, and ${9/2}^+$ states in all four isotopes. These sequences were interpreted as arising from the weak coupling of the proton quasiparticle with the core states in even-even Zn isotopes, suggesting the potential observation of mixed configurations due to broken proton and neutron pairs at higher excitation energy \cite{Stefanescu2009}. Concurrently, Diriken \textit{et al.}~\cite{Diriken2010} conducted Coulomb excitation experiments to investigate collectivity in $^{71,73}$Ga, which revealed a structural change induced by the $g_{9/2}$ orbital as the neutron number increased from $N=40$ to $N=42$.  Thus, the addition of $g_{9/2}$ neutrons drives the nuclear shape toward larger deformation, resulting in more pronounced rotational-like structures built on single-particle proton levels.

The present study investigates the structure of $N=38$ $^{69}$Ga. Low-lying levels in this nucleus have been explored experimentally in previous studies. The earliest investigation employed Coulomb excitation with $\alpha$-particles \cite{Fagg1956}. Although the reduced transition probability of the ${1/2}^- \rightarrow {3/2}^-$ ground-state transition involving the $p_{3/2}$ and $p_{1/2}$ orbitals was reported, no other levels were observed due to low statistics and the modest detector system used.
Subsequent experimental efforts included electron-capture decay studies \cite{temperley1965,ZOLLER1969} and other reactions such as $^{69}$Ga($\gamma, \gamma')$ \cite{LANGHOFF1968, moreh1973}, $^{69}$Ga($n, n'\gamma)$ \cite{Velkley1969}, $^{68}$Zn$(d,n\gamma)$ \cite{Couch1970}, $^{69}$Ga$(\alpha,\alpha'\gamma)$ and $^{66}$Zn$(\alpha,p\gamma)$ \cite{IVASCU1974}, $^{68}$Zn$(^{3}He,d\gamma)$ \cite{RICCATO1974}, $^{66}$Zn$(\alpha,p\gamma)$ \cite{Harms-Ringdahl1974}, $^{69}$Ga($p, p'\gamma)$ \cite{Paradellis1978}, $^{68}$Zn$(p,\gamma)$ \cite{Paradellis1981}, and $^{64}$Ni$(^7Li, 2n\gamma)$ \cite{Bakoyeorgos1982}. These investigations reported negative-parity states mediated by single-particle excitations and core-coupling configurations involving the $\pi(p_{3/2}, f_{5/2}, p_{1/2}, f_{7/2}$) orbitals.

 The most recent published study, performed by Ref. \cite{Bakoyeorgos1982}, utilized the $^{64}$Ni$(^{7}\mathrm{Li}, 2n\gamma)$ reaction. The experimental setup consisted of two Ge(Li) detectors positioned at 0\degree and 90\degree relative to the beam direction. A level scheme based on coincidence relationships was developed up to 4.5 MeV excitation energy. Spin-parity assignments were proposed based on angular distribution and directional correlation from oriented states measurements. Due to limitations in the experimental system and limited detection statistics, spin and parity assignments were firmly established only for states up to approximately 3 MeV. Nevertheless, two distinct band-like structures of positive and negative parity were delineated up to ${17/2}^+$ and ${9/2}^-$, respectively. The authors suggested that the spin-parities in these bands could be interpreted within the framework of particle-core coupling, specifically $\pi(g_{9/2} \otimes J)$ and $\pi(f_{5/2} \otimes J)$ configurations coupled to $J=0^+, 2^+, 4^+, 6^+$ quadrupole vibrations for positive- and negative-parity bands, respectively\cite{Bakoyeorgos1982}. 
 
In the present study, medium- and high-spin states in $^{69}$Ga have been investigated via the fusion-evaporation reaction $^{26}$Mg($^{48}$Ca, $p4n$) performed at the Argonne National Laboratory. The emitted $\gamma$ rays were detected using the Gammasphere array \cite{I.Y.Lee1990NPA} in conjunction with the Fragment Mass Analyzer \cite{DAVIDS19891224}. The low-lying level structure of $^{69}$Ga has been significantly expanded up to $\sim15$ MeV based on $E_\gamma$-$E_\gamma$-$E_\gamma$ coincidence measurements. The multipolarities of the emitted $\gamma$ transitions were determined by angular distribution and correlation of the emitted $\gamma$ rays measurements, thereby enabling firm spin assignments to the excited states.
This study confirms most of the low-lying levels and extends the negative-parity structures reported in Ref.~\cite{Bakoyeorgos1982, sharma2025aspectssingleparticleexcitations}. In addition, three collective, positive-parity, high-spin sequences with quadrupole transitions were observed in coincidence with the low-spin states. The experimental low-lying levels, are compared with the results of shell-model calculations performed using the jj44b and JUN45 interractions. The collective sequences observed at high spins were compared with calculations carried out within the framework of the tilted-axis-cranking covariant density functional theory (TAC-CDFT). Details about the experimental technique and data analysis are given in Section \ref{exp}. The present level scheme is introduced in Section~\ref{levelscheme}. The interpretation of experimental and theoretical results is given in Section \ref{discussion}, and Section \ref{summary} presents the conclusions of the present work.


\section{EXPERIMENT}\label{exp}
Intermediate- and high-spin states in the odd-even $^{69}$Ga nucleus were populated following the $^{26}$Mg($^{48}$Ca, $p4n$) reaction, performed in inverse kinematics at Argonne National Laboratory (ANL). The measurement utilized a highly-enriched $(>98 \%)$, self-supporting $^{26}$Mg target with a thickness of $0.5\, \mathrm{mg/cm^2}$, bombarded by a 195-MeV $^{48}$Ca beam from the Argonne Tandem Linac Accelerator System (ATLAS). The $p4n\gamma$ reaction channel that results in the production of $^{69}$Ga ions accounted for approximately 27\% of the total reaction cross section.
The emitted $\gamma$ rays from the de-excitation process were detected using Gammasphere \cite{I.Y.Lee1990NPA}, which consisted of 78 high-purity germanium (HPGe) detectors with BGO Compton suppressors at the time of the experiment. 

Particle detectors were incorporated into the experimental setup alongside Gammasphere to facilitate clean selection of reaction exit channels. Neutrons were measured using a Neutron Shell made of liquid scintillation detectors \cite{SARANTITES2004473} which replaced the HPGe-BGO modules in the five most forward rings of Gammasphere with respect to the beam direction. Charged-particle detection and identification was carried out using Microball~\cite{SARANTITES1996418} along with two double-sided silicon strip detectors (DSSD) positioned within the target chamber \cite{yulia}. The recoils were transported through the Fragment Mass Analyzer (FMA) and separated at the focal plane based on their mass-to-charge ratio ($A/Q$). A parallel plate avalanche counter (PPAC), positioned at the focal plane, was used to record the position and time-of-flight information of the recoils on an event-by-event basis. For $Z$ identification, a three-fold segmented ionization chamber, located behind the focal plane was used to measure the energy loss of the reaction products. The events were recorded under the condition that detection of a reaction product in the PPAC detector be in coincidence with at least two $\gamma$ rays detected by Gammasphere within a prompt time window. 

\begin{figure}[ht]
    \centering
    \includegraphics[width=\columnwidth]{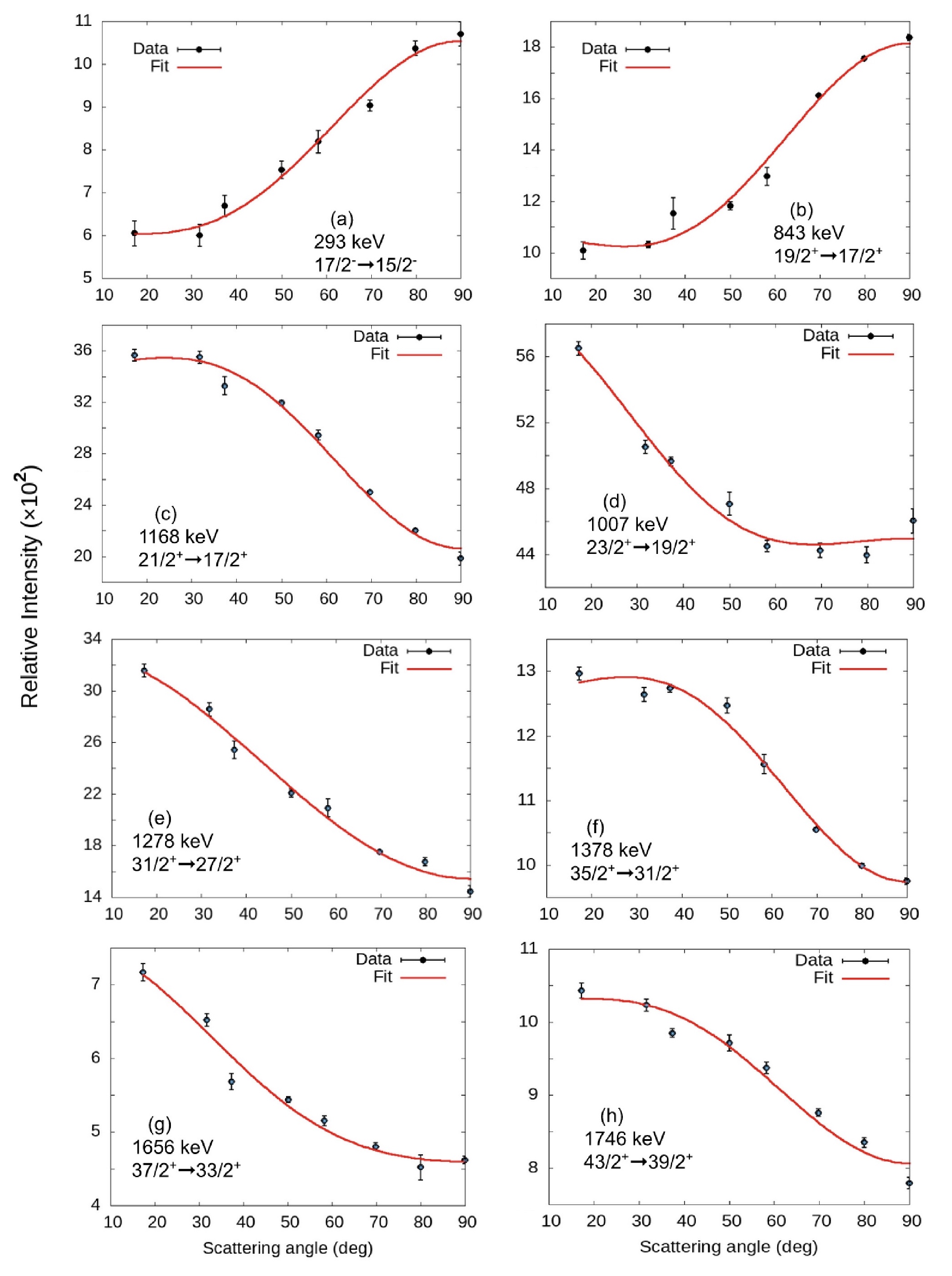}
    \caption{Angular distributions for some transitions in $^{69}$Ga. Experimental data are shown as black circles, while the angular-distribution fits correspond to the red curves.}
    \label{fig:angdis}
\end{figure}

Due to the high yield of $^{69}$Ga ions, information from Microball, DSSD and the Neutron Shell was not included in the present study. Only Compton-suppressed $\gamma$-$\gamma$ and $\gamma$-$\gamma$-$\gamma$ coincidence events from Gammasphere array were considered relevant for the analysis, once the coincidence conditions on the $^{69}$Ga recoil identification were fulfilled. The events associated with the de-excitation of Ga ions populated during the reaction were sorted into a fully symmetrized three-dimensional $E_\gamma$-$E_\gamma$-$E_\gamma$ coincidence cube, while events related to $^{69}$Ga were sorted into fully symmetrized two-dimensional $E_\gamma$-$E_\gamma$ coincidence matrices. Subsequent data analysis was performed using the RADWARE analysis package \cite{RADFORD1995297}. The level scheme for the $^{69}$Ga nucleus was constructed based on coincidence relationships and intensity considerations. 

Spin and parity assignments for the newly identified levels were proposed by examining the angular distribution of the emitted $\gamma$ rays. 
This was carried out using $E_\gamma$-$E_\gamma$ coincidence matrices sorted such that the energies of $\gamma$ rays detected at specific Gammasphere angles, measured with respect to the beam direction, $E_\gamma(\theta)$, were incremented on one axis, and those measured at all angles placed on the other $E_\gamma$(any) one. Since angular distributions are symmetric about $90^\circ$, statistics were improved by combining adjacent rings of Gammasphere and those corresponding to angles symmetric with respect to the beam axis in the forward and backward hemispheres, forming a total of eight matrices at average angle $17.3^\circ$, $31.7^\circ$, $37.4^\circ$, $50.1^\circ$, $58.3^\circ$, $69.8^\circ$, $79.9^\circ$, and $90.0^\circ$. By placing a gate on the $E_{\gamma}$(any) axis and projecting on the respective $E_{\gamma}(\theta)$ one, background-subtracted and efficiency-corrected spectra were generated from which $\gamma$-ray intensities for transitions of interest were extracted at each angle and fitted to the angular distribution function, 
$W(\theta)=a_0[1+a_2P_2(\cos\theta)+a_4P_4(\cos\theta)]$, where $P_2$ and $P_4$ are Legendre polynomials. Representative fits for the angular distribution of selected transitions are displayed in Fig.~\ref{fig:angdis}.

For transitions observed with low statistics, an angular correlation ratio of $\gamma$-ray intensities observed in detectors at backward angles to those seen around $90^\circ$ was derived. Two coincidence matrices were created for this purpose. In one, $E_\gamma$(b)-vs-$E_\gamma$(any), $\gamma$ rays observed in detectors at backward angles ($142.6\degree$, $148.3\degree$ and $162.7\degree$) were placed along one axis, while those observed at any angle were incremented along the other. The other matrix, $E_\gamma(\sim90\degree)$-vs-$E_\gamma(\textrm{any})$, was similarly constructed but with coincidence transitions observed in detectors positioned close to $90^\circ$ incremented on the one axis. A correlation ratio, defined by $R_{\textrm{ac}}={I_\gamma(\theta_b,\textrm{any})}/{I_\gamma(\theta_{\sim90\degree},\text{any})}$, where $I_\gamma(\theta_x,\textrm{any})$ is the relevant $\gamma$-ray intensity, was obtained by placing gates on the corresponding $E_\gamma(\textrm{any})$ axes in the two matrices. This ratio, which is independent of the multipolarity of the gating transition, was established to be greater than 1.2 for stretched-quadrupole transitions and less than 0.8 for stretched-dipole ones~\cite{rudolph199956Ni,ayangeakaa2022s, sensharma2022s}. Together with the angular distribution, the $R_{\textrm{ac}}$ values were used to distinguish between dipole and quadrupole transitions. The energies, relative intensities, associated angular distribution coefficients, $R_{ac}$ ratios, as well as the multipolarity assignments for all observed transitions in the present study, are given in Table~\ref{table1}.

\LTcapwidth=\textwidth 
\setlength\tabcolsep{6pt}
{\renewcommand{\arraystretch}{1.0}%
\begin{longtable*}{@{\extracolsep{\fill}} cccccccc @{\extracolsep{\fill}}}
\caption{\label{table1}
$\gamma$-ray energies, relative intensities, energy of the initial state, initial and final spins between which the transition occurs, angular correlation ratios ($R_\text{ac}$), experimental angular distribution coefficients ($A_2$) and ($A_4$), and assigned multipolarities for the transitions shown in Fig. ~\ref{fg:low_sch}. The intensities ($I_\gamma$) of the $\gamma$-ray transitions were corrected for detector efficiency and normalized to the 1337.3-keV ground-state transition. Parity assignments for newly established states followed the previously established level scheme of Ref. \cite{Bakoyeorgos1982}, by assuming that the transitions connect levels with the same parity. All multipolarities enclosed in parentheses are uncertain. }
\\
\hline \hline
$E_{\gamma}$ (keV) & $I_\gamma$ & $E_i$ (keV) & $I_\text{i}^\pi$ $\to$ $I_\text{f}^\pi$ & $A_2$ & $A_4$ & $R_\text{ac}$ & Mult. \\ \hline
\endfirsthead
\multicolumn{8}{c}%
{\tablename\ \thetable\ -- \textit{Continued from previous page.}} \\
\hline
$E_{\gamma}$ (keV) & $I_\gamma$ & $E_i$ (keV) & $I_\text{i}^\pi$ $\to$ $I_\text{f}^\pi$  & $A_2$ & $A_4$ & $R_\text{ac}$& Mult. \\ \hline
\endhead
\hline
\endfoot
\endlastfoot

  153.3(3)     	  & 3.5(7)       &  3543.0(5)        &$   15/2^{+}  \rightarrow 15/2^{(+)} $&   -0.48(2)   &   0.23(3) 	  & 0.62(4)       &    $(M)1$      \\
  208.0(3)     	  & 4.1(8)       &  1972.4(24)       &$    9/2^{+}  \rightarrow    9/2^{-} $&   -          &   - 			  & 0.80(18)      &       $E1$         \\
  248.4(5)     	  & 4.7(6)       &  5982.6(6)        &$   25/2^{+}  \rightarrow   25/2^{-} $&   -          &   - 			  & 0.66(9)       &       $E1$\\
  293.0(4)     	  & 2.8(3)       &  4078.5(6)        &$   17/2^{-}  \rightarrow   15/2^{-} $&   -0.42(7)   &   0.12(10) 	  & 0.65(2)        &       $M1+E2$      \\
  341.2(4)     	  & 3.2(6)       &  3976.0(8)        &$ 19/2^{(+)}  \rightarrow   17/2^{+} $&   -          &   -             & 0.55(14)       &       $(M)1$      \\
  380.9(2)     	  & 6.4(5)       &  1487.9(3)        &$    7/2^{-}  \rightarrow    5/2^{-} $&   -          &   -             & 0.66(5)        &       $M1+E2$      \\
  399.4(7)     	  & 3.5(6)       &  4477.9(5)        &$ 19/2^{(+)}  \rightarrow   17/2^{-} $&   -0.38(2)   &   -0.0           & 0.74(7)        &       $(E)1$         \\      
  448.3(4)     	  & 41.8(9)      &  4526.8(7)        &$   21/2^{-}  \rightarrow   17/2^{-} $&   0.17(2)    &   -0.19(6) 	  & 1.15(2)        &       $E2$         \\      
  484.5(5)     	  & 22.9(8)      &  1972.4(24)       &$    9/2^{+}  \rightarrow    7/2^{-} $&   -0.37(3)   &   0.06(5) 	  & 0.70(2)        &       $E1$         \\      
  497.3(6)     	  & 14.0(12)     &  5982.6(6)        &$   25/2^{+}  \rightarrow 23/2^{(+)} $&   -0.13(9)   &   -0.15(13) 	  & 0.80(3)        &       $M1+E2$      \\
  535.6(5)     	  & 1.9(18)      &  5485.2(5)        &$ 23/2^{(+)}  \rightarrow (21/2^{+}) $&   -          &   -  			  & -              &       $(M1+E2)$    \\
  543.0(10)    	  & 2.3(15)      &  3785.6(5)        &$   15/2^{-}  \rightarrow   13/2^{-} $&   -          &   -  			  & -              & 	   $(M1+E2)$    \\
  546.4(5)     	  & 2.8(14)      &  4949.5(8)        &$ (21/2^{+})  \rightarrow (19/2^{+}) $&   -          &   -  			  & 0.66(32)       &       $M1+E2$    \\
  574.5(3)     	  & 100(20)      &  574.48(23)       &$    5/2^{-}  \rightarrow    3/2^{-} $&   -0.35(7)   &   0.13(10)      & 0.71(1)        &       $M1+E2$      \\
  602.3(8)     	  & 2.7(6)       &  3321.1(5)        &          -   $\rightarrow   13/2^{+} $&   -          &   -             & -              &       $(M1+E2)$     \\
  635.1(7)       	  & 76.5(4)      &  1972.4(24)       &$    9/2^{+}  \rightarrow    7/2^{-} $&   -0.21(3)   &   0.13(4) 	  & 0.93(2)        &       $E1$         \\
  637.3(6)     	  & 6.6(9)       &  4179.8(7)        &$ (19/2^{+})  \rightarrow   15/2^{+} $&   -          &   - 			  & 1.27(41)       &       $E2$         \\
  671.0(4)     	  & 7.3(7)       &  3389.7(5)        &$ 15/2^{(+)}  \rightarrow   13/2^{+} $&   -0.09(8)   &   -0.10(11) 	  & 0.69(2)        &       $M1+E2$      \\
  694.5(2)     	  & 3.6(5)       &  3413.3(10)       &$ 15/2^{(+)}  \rightarrow   13/2^{+} $&   -          &   - 			  & 0.81(5)        &       $M1+E2$      \\
  746.2(7)     	  & 96.9(24)     &  2718.8(5)        &$   13/2^{+}  \rightarrow    9/2^{+} $&   0.11(1)    &   -0.06(1) 	  & 1.21(2)        &       $E2$         \\
  748.3(11)    	  & 4.5(7)       &  4929.0(9)        &$ (21/2^{+})  \rightarrow (19/2^{+}) $&   -          &   -             & 0.85(7)        &       $M1+E2$      \\
  762.8(2)     	  & 4.1(2)       &  1337.30(25)      &$    7/2^{-}  \rightarrow    5/2^{-} $&   -          &   -             & 0.60(3)        &       $M1+E2$      \\
  774.2(2)     	  & 1.8(6)       &  5485.2(5)        &$ 23/2^{(+)}  \rightarrow (19/2^{+}) $&   -          &   -             & -              &       $(E2)$       \\
  791.5(9)     	  & 3.6(5)       &  7825.9(12)       &             -                       &   -          &   -             &    -   &             $(M1+E2)$    \\
  824.3(2)     	  & 9.1(6)       &  3543.0(5)        &$   15/2^{+}  \rightarrow   13/2^{+} $&   -          &   -             & 0.76(7)        &       $M1+E2$      \\
  835.9(8)     	  & 50.9(10)     &  4078.5(6)        &$   17/2^{-}  \rightarrow   13/2^{-} $&   -0.14(2)   &   0.01(2)       & 0.84(1)        &       $M1+E2$      \\
  843.1(2)     	  & 30.8(11)     &  4477.9(5)        &$ 19/2^{(+)}  \rightarrow   17/2^{+} $&   -0.44(20)  &   0.03(4)       & 0.63(4)        &       $M1+E2$      \\
  913.4(3)     	  & 14.7(10)     &  1487.9(3)        &$    7/2^{-}  \rightarrow    5/2^{-} $&   -          &   -             & 0.78(2)        &       $M1+E2$      \\
  916.1(3)     	  & 61.8(17)     &  3634.8(5)        &$   17/2^{+}  \rightarrow   13/2^{+} $&   0.30(2)    &   0.23(3)       & 1.05(2)        &       $E2$         \\
  956.3(4)     	  & 3.8(7)       &  5034.8(8)        &$ 21/2^{(-)}  \rightarrow   17/2^{-} $&   -          &   -             & 1.45(7)        &       $(E2)$       \\
  986.5(6)     	  & 5.2(13)      &  5513.3(9)        &$ 23/2^{(-)}  \rightarrow   21/2^{-} $&   -          &   -             & 0.46(1)        &       $M1+E2$       \\
  1004.0(8)    	  & 8.2(26)      &  3722.7(6)        &$   17/2^{+}  \rightarrow   13/2^{+} $&   -          &   -             & 1.04(2)        &       $E2$         \\
  1007.2(5)    	  & 11.7(12)     &  5485.2(5)        &$ 23/2^{(+)}  \rightarrow 19/2^{(+)} $&   0.16(1)    &   0.09(2)       & 0.94(5)        &       $E2$         \\
  1013.4(3)    	  & 3.5(11)      &  4403.1(6)        &$ (19/2^{+})  \rightarrow 15/2^{(+)} $&   -          &   -             & 1.02(29)       &       $E2$       \\
  1066.9(7)    	  & 3.6(9)       &  3785.6(5)        &$   15/2^{-}  \rightarrow   13/2^{+} $&   -          &   -             &    -           &       $(E1)$       \\
  1080.5(2)    	  & 7.6(14)      &  4803.2(6)        &$   21/2^{+}  \rightarrow   17/2^{+} $&   -          &   -             & -       &       $(E2)$       \\
  1107.0(8)    	  & 12.6(12)     &  1107.0(8)        &$    5/2^{-}  \rightarrow    3/2^{-}  $&   -          &   -             & 0.35(12)       &       $M1+E2$      \\
  1117.2(6)    	  & 20.3(13)     &  3785.6(5)        &$   15/2^{-}  \rightarrow   11/2^{-}  $&   -          &   -             & 1.03(4)        &       $E2$         \\
  1125.0(2)    	  & 29.1(9)     &  7107.6(7)         &$   29/2^{+}  \rightarrow   25/2^{+}  $&   0.13(4)    &   -0.10(5)      & 1.17(22)       &       $E2$         \\
  1150.8(2)    	  & 6.4(7)      &  5677.8(7)         &         -    $\rightarrow   21/2^{-}  $& -          &   -             & 1.19(7)        &       $E2$         \\
  1168.4(8)         & 25.6(15)     &  4803.2(6)        &$  21/2^{+}  \rightarrow   17/2^{+}   $&   0.41(2)    &   -0.17(3)      &   1.21(9)      &       $E2$         \\
  1179.2(2)    	  & 15.3(14)     &  5982.6(6)        &$   25/2^{+}  \rightarrow   21/2^{+} $&-           &   -             & 1.20(76)       &       $E2$         \\
  1180.6(7)         & 10.4(22)     &  2668.5(4)      &$   11/2^{-}  \rightarrow    7/2^{-} $&  -           &   -             & 1.46(67)       &       $E2$         \\
  1189.9(5)    	  & 67.1(5)      &  1764.4(5)        &$    9/2^{-}  \rightarrow    5/2^{-} $&-           &   -             & 1.01(28)       &       $E2$         \\
  1207.2(10)     	  & 30.0(10)     &  5734.2(7)    &$   25/2^{-}  \rightarrow   21/2^{-} $&   0.40(8)    &   -0.10(11)     & 1.16(23)       &       $E2$         \\
  1212.4(3)         & 5.7(2)       &  2319.4(5)      &$      (7/2)  \rightarrow    5/2^{-} $&  -           &   -             & 0.43(6)        &       $M1+E2$      \\
  1221.0(3)    	  & 4.3(8)       &  6734.4(8)        &$ (27/2^{-})  \rightarrow 23/2^{(-)} $&-           &   -             & 1.44(6)        &       $(E2)$       \\
  1235.4(4)    	  & 16.4(9)      &  8342.8(9)        &$   33/2^{+}  \rightarrow   29/2^{+} $&0.17(1)     &   0.14(1)       & 1.18(3)        &       $E2$         \\
  1253.2(6)    	  & 2.3(12)      &  7235.7(8)        &$ 27/2^{(+)}  \rightarrow   25/2^{+} $&-           &   -             & -       &       $(M1+E2)$    \\
  1259.6(2)    	  & 17.1(13)      &  5737.5(6)       &$ 23/2^{(+)}  \rightarrow 19/2^{(+)} $&-           &   -             & 1.06(34)       &       $E2$         \\
  1277.5(5)    	  & 17.6(13)     &  6080.7(8)        &$ 25/2^{(+)}  \rightarrow   21/2^{+} $&0.55(3)     &   0.01(4)       & 1.10(17)       &       $E2$         \\
  1278.3(3)    	  & 13.95(24)     &  8513.8(7)       &$ 31/2^{(+)}  \rightarrow 27/2^{(+)} $&-           &   -             & 1.14(3)        &       $E2$         \\
  1297.8(10)        & 2.5(8)       &  4711.0(8)      &$ (19/2^{+})  \rightarrow 15/2^{(+)} $&  -           &   -             & 1.69(16)       &       $(E2)$       \\
  1327.3(6)    	  & 12.7(13)     &  7061.3(9)        &$   29/2^{-}  \rightarrow   25/2^{-} $&0.25(3)     &   -0.20(4)      & 1.26(4)        &       $E2$         \\
  1331.2(6)         & 16.9(18)     &  2668.5(4)      &$   11/2^{-}  \rightarrow    7/2^{-} $&  -           &   -             & 1.71(89)       &       $(E2)$       \\
  1337.3(12)        & 100.0(8)     &  1337.30(25)    &$    7/2^{-}  \rightarrow    3/2^{-} $&  -           &   -             & 1.29(9)        &       $E2$         \\
  1354.3(1)         & 5.6(11)      &  3118.7(6)      &$   11/2^{-}  \rightarrow    9/2^{-} $&  -           &   -             & 1.53(8)        &       $(E2)$       \\
  1356.6(4)         & 3.6(4)       &  7034.4(8)      &             -           &-           &   -             & -       &       $(E2)$       \\
  1378.2(9)         & 9.30(10)     &  9892.1(12)     &$ 35/2^{(+)}  \rightarrow 31/2^{(+)} $&  0.21(1)     &   -0.10(2)      & 1.68(9)        &       $E2$         \\
  1397.9(5)         & 6.6(2)       &  1972.4(24)     &$    9/2^{+}  \rightarrow    5/2^{-} $&  -           &   -             & 0.66(44)       &       $M2$         \\
  1406.4(8)         & 4.3(11)      &  8513.8(7)      &$ 31/2^{(+)}  \rightarrow   29/2^{+} $&  -           &   -             & 0.86(19)       &       $M1+E2$      \\
  1421.2(2)         & 13.64(21)      &  7502.1(8)    &$ 29/2^{(+)}  \rightarrow 25/2^{(+)} $&  0.48(5)    &   0.02(7)       & 1.08(4)        &       $E2$         \\
  1478.3(8)         & 58.2(13)     &  3242.6(7)      &$   13/2^{-}  \rightarrow    9/2^{-} $&  -           &   -             & 1.09(3)        &       $E2$         \\
  1487.8(4)        & 24.4(13)     &  1487.9(3)       &$    7/2^{-}  \rightarrow    3/2^{-} $&  -           &   -             & 1.11(6)        &       $E2$         \\
  1498.0(5)        & 15.50(95)       &  7235.7(8)    &$ 27/2^{(+)}  \rightarrow 23/2^{(+)} $&  -           &   -             & 1.33(34)       &       $E2$         \\
  1509.2(2)        & 3.20(11)     &  5485.2(5)       &$ 23/2^{(+)}  \rightarrow 19/2^{(+)} $&  -           &   -             & 1.20(6)        &       $E2$         \\
  1510.2(2)        & 10.9(8)     &  8571.5(8)        &$   33/2^{-}  \rightarrow   29/2^{-} $&  0.50(16)    &   -0.34(21)     & 1.09(15)       &       $E2$         \\
  1553.5(6)        & 7.8(9)     &  11445.5(14)       &$ 39/2^{(+)}  \rightarrow 35/2^{(+)} $&  0.61(11)    &   -0.13(14)     & 1.23(14)       &       $E2$         \\
  1594.6(9)        & 4.5(10)      &  11593.3(13)     &$   41/2^{+}  \rightarrow   37/2^{+} $&  -           &   -             & 1.11(3)        &       $E2$         \\
  1614.5(3)        & 6.4(13)     &  9116.6(8)        &$ 33/2^{(+)}  \rightarrow 29/2^{(+)} $&  -           &   -             & 1.34(10)       &       $E2$         \\
  1619.8(8)        & 2.8(7)       &  10191.4(9)      &$   35/2^{-}  \rightarrow   33/2^{-} $&  -           &   -             & -       &       $(E2)$       \\
  1655.7(3)        & 12.7(14)     &  9998.5(9)       &$   37/2^{+}  \rightarrow   33/2^{+} $&  0.34(3)     &   0.08(4)       & 1.02(10)       &       $E2$         \\
  1746.2(6)        & 6.2(13)      &  13191.9(15)     &$ 43/2^{(+)}  \rightarrow 39/2^{(+)} $&  0.19(2)     &   -0.07(2)      & 1.26(9)        &       $E2$         \\
  1773.0(20)       & 5.8(13)      &  7507.0(21)      &$ 29/2^{(-)}  \rightarrow   25/2^{-} $&  -           &   -             & 1.10(6)        &       $E2$         \\
  1797.6(9)        & 3.6(13)     &  10914.2(12)      &$ 37/2^{(+)}  \rightarrow 33/2^{(+)} $&  0.12(4)     &   -0.01(6)      & 1.21(4)        &       $E2$         \\
  2009.6(7)        & 3.3(7)       &  15201.3(16)     &$ 47/2^{(+)}  \rightarrow 43/2^{(+)} $&  -           &   -             & 1.43(9)        &       $E2$       \\
  2098.8(4)        & 2.81(25)      &  10670.4(10)    &$   37/2^{-}  \rightarrow   33/2^{-} $&  -           &   -             & 1.47(13)       &       $E2$       \\
  2115.5(3)        & 1.82(25)      &  13029.6(13)    &$ 41/2^{(+)}  \rightarrow 37/2^{(+)} $&  -           &   -             & 1.25(8)        &       $E2$          \\
  2186.9(10)       & 2.6(7)       &  9694.0(21)      &$ 33/2^{(-)}  \rightarrow 29/2^{(-)} $&  -           &   -             & 1.05(27)       &       $E2$          \\
  2223.5(7)        & ($<1$)         &  15253.3(15)     &$ (45/2^{+})  \rightarrow 41/2^{(+)} $&  -           &   -             & 1.14(86)       &       $E2$          \\
  2689.2(2)        & ($<1$)         &  14282.4(13)     &$ (45/2^{+})  \rightarrow   41/2^{+} $&  -           &   -             & 1.26(59)       &       $E2$          \\
\hline \hline
\end{longtable*}}

\section{LEVEL SCHEME}\label{levelscheme}

\begin{figure*}[ht]
	\centering
    \includegraphics[width=1.05\textwidth]{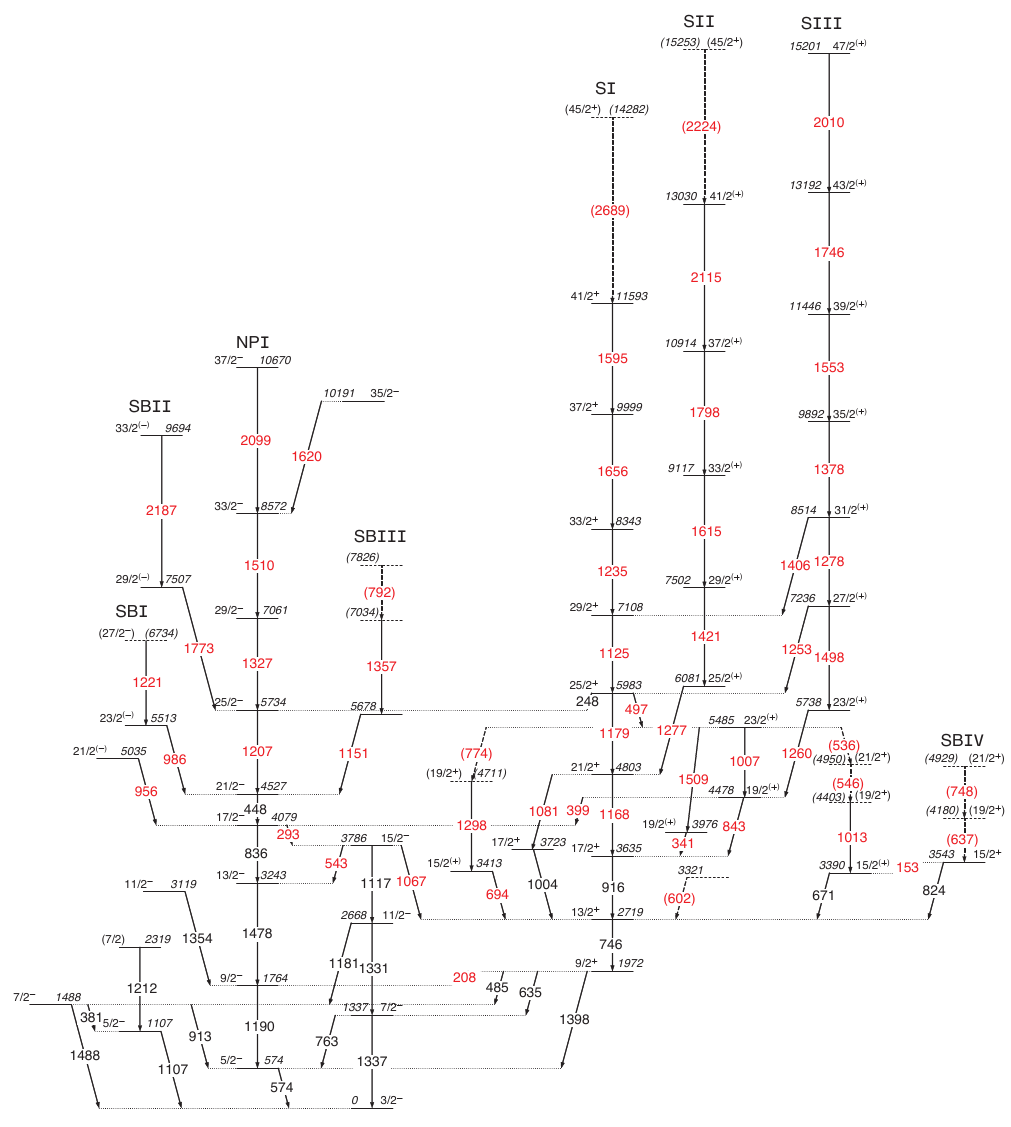}
	\caption{Level scheme of $^{69}$Ga. 
    Previously known transitions are shown in black, newly-placed ones are in red. Tentatively placed transitions are enclosed in parentheses.}
	\label{fg:low_sch}
\end{figure*}

The partial $^{69}$Ga level scheme, established in this work is presented in Fig. \ref{fg:low_sch}. The placement of $\gamma$ rays in the level scheme is determined through a combination of $\gamma-\gamma-\gamma$ coincidence relationships, relative intensity balances, and energy sums. The spin and parity assignments are based on angular distribution and correlation ratios of the emitted $\gamma$ rays. These assignments for newly placed levels also rely on those proposed for lower-lying states in previous works \cite{Paradellis1978,Paradellis1981,Bakoyeorgos1982}. Prior studies \cite{Almar1972,Harms-Ringdahl1974,Bakoyeorgos1982, Paradellis1978,Paradellis1981,Bakoyeorgos1982} established a ground-state ${3/2}^-$ spin-parity, consistent with the presence of three valence protons in the $p_{3/2}$ orbital. The ${5/2}^-$ levels at 574 and 1107 keV as well as the ${7/2}^-$ states at 1337 and 1488 keV were previously identified in a ($p,\gamma$) reaction \cite{Paradellis1981}, where the negative parity was based on the measured angular distributions and on lifetime measurements. These assignments were also confirmed by Bakoyeorgos \textit{et al.} \cite{Bakoyeorgos1982} using the $^{64}$Ni + $^{7}$Li reaction. The present study validates the previously established structure and significantly extends the level scheme toward higher spins; for example, up to $\sim15$ MeV. 
All identified $\gamma$ rays and their associated properties are summarized in Table~\ref{table1}.

\begin{figure}[ht]
    \centering
    \includegraphics[width=\columnwidth]{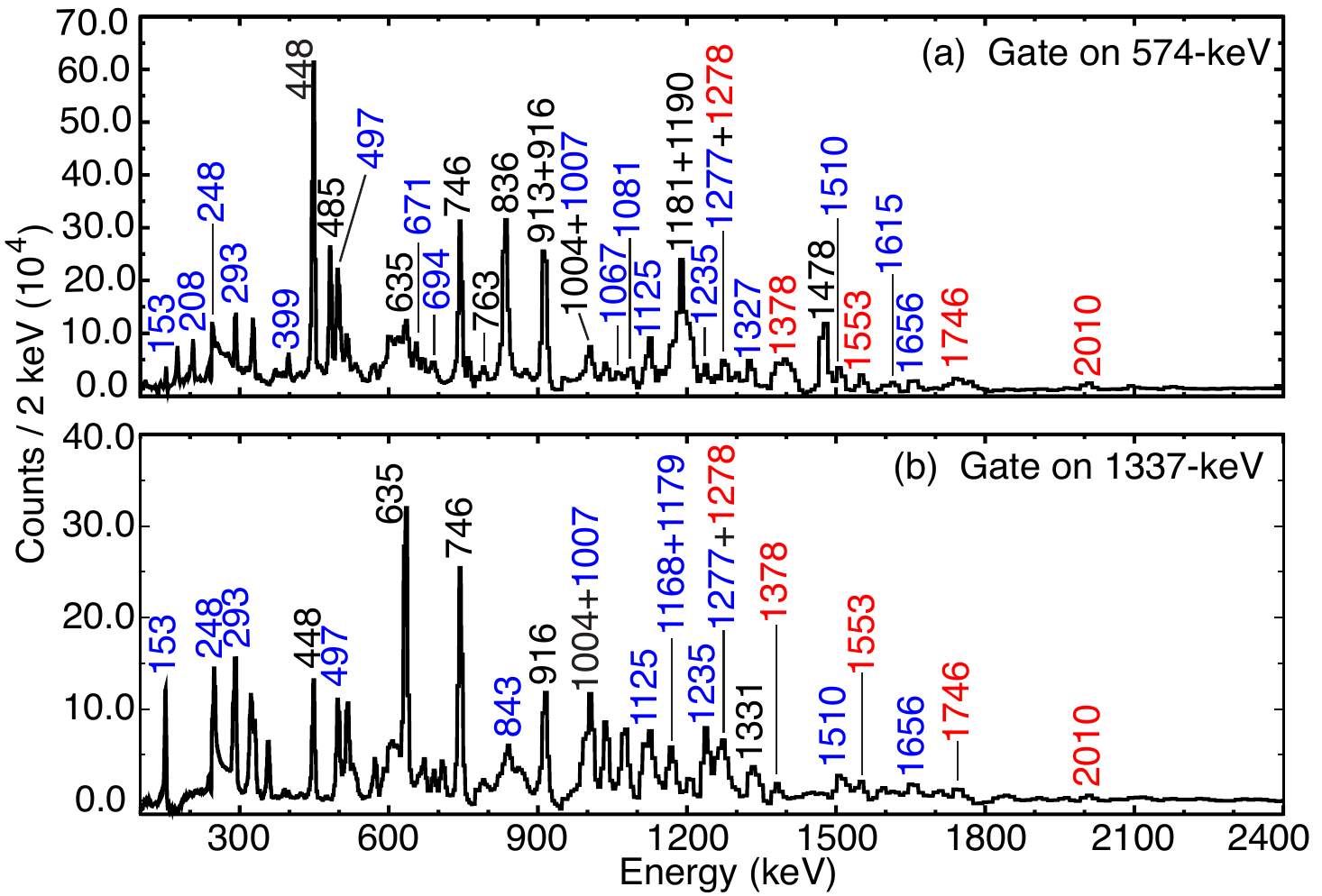}
    \caption{Coincidence spectra resulting from a single gate on the 574- and 1337-keV transitions. The black-, blue- and red-colored energies are the previously known, newly identified and the rotational-like quadrupole transitions, respectively. See text for details.}
    \label{fig:g574_1337}
\end{figure}


Figures \ref{fig:g574_1337} (a) and (b) present, respectively, background-subtracted coincidence spectra obtained by placing single gates on the 574- and 1337-keV ground-state transitions. These spectra identify some of the dominant transitions feeding the ${3/2}^-$ ground state with peaks marked in black, blue, and red corresponding to the previously known, newly-identified low- and high-spin levels, respectively. The highly-intense 1190-1478-836-448-keV $\gamma$-ray cascade - directly populating the 574-keV level is clearly visible in Fig. \ref{fig:g574_1337} (a). The present study confirms the presence of a 763-keV, ${7/2}^-\rightarrow {5/2}^-$ $\gamma$ ray depopulating the level at 1337 keV, and a 913-keV, ${7/2}^-\rightarrow {5/2}^-$ transition connecting the 1488- and 574-keV states, as reported in Refs.~\cite{Bakoyeorgos1982,Paradellis1981}. This 1488-keV level has also been confirmed to decay to the ground state via a 1488-keV, ${7/2}^-\rightarrow {3/2}^-$ transition, and to the level at 1107 keV via a 381-keV, ${7/2}^-\rightarrow {5/2}^-$ $\gamma$ ray. Additionally, Fig. \ref{fig:g574_1337} (a) presents the 746-, 916-, and 1004-keV transitions in the low-spin region of sequence S I. These transitions are in coincidence with the 574-keV one via the 913-485-1190-208-keV, and the 763-635-keV cascades. In Fig. \ref{fig:g574_1337} (b), the previously known low-spin 635-, 746-, and 916-keV transitions atop the 1337-keV level are prominent. Newly identified transitions connecting medium-spin states are also observed, including the 497-, 843-, 1007-, 1081-, 1125-, 1168-, 1179-, and 1235-keV $\gamma$ rays. In addition, the newly identified high-spin sequence labelled S III, comprising the 1278-, 1378-, 1553-, 1746-, and 2010-keV quadrupole transitions (marked in red), is observed in coincidence with the gating 1337-keV $\gamma$ ray. Other newly identified transitions, such as the 1207-, 1327-, and 1510-keV ones, connecting the medium-to-high spin levels in the negative-parity sequence, NP I and depopulating via the 293-1117-1331-keV cascade, are also seen in coincidence with the 1337-keV transition.

\begin{figure}[ht]
    \centering
    \includegraphics[width=\columnwidth]{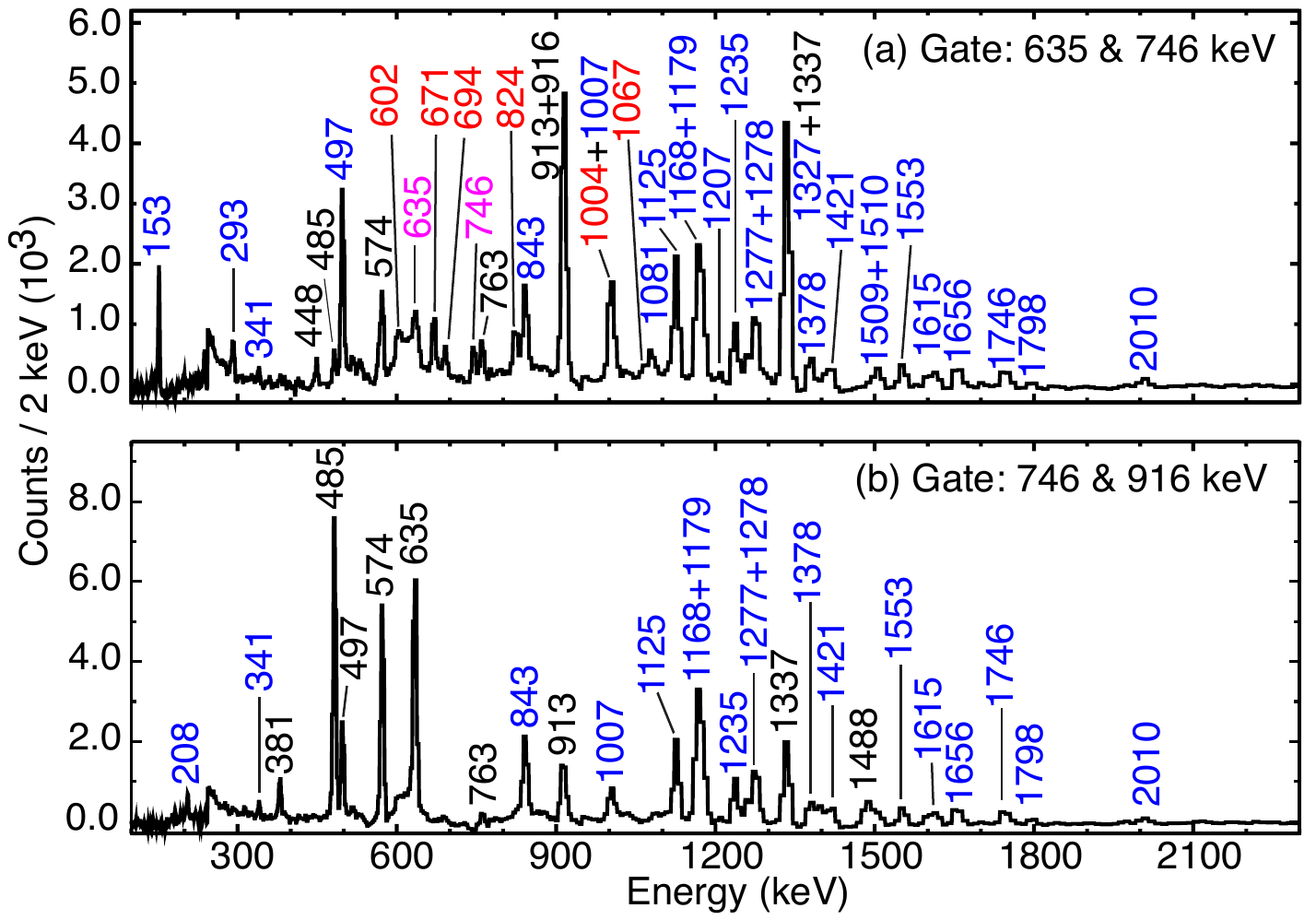}
    \caption{Coincidence spectra resulting from double gates on (a) 635- and 746-keV and (b)746- and 916-keV $\gamma$-ray transitions. The previously known transitions are marked in black, while the colored $\gamma$ rays are the newly established ones. Transitions populating the 2719-keV state are marked in red. The doublet $\gamma$ rays in coincidence with the gated transitions are in magenta.}
    \label{fig:g636_747_916}
\end{figure}

The angular distribution and $R_{ac}$ ratios presented in Table \ref{table1} confirmed the spin-parity assignments of ${7/2}^-$, ${9/2}^+$, ${13/2}^+$, and ${17/2}^+$ to the 1337-, 1972-, 2719-, and 3635-keV levels, respectively, as previously proposed by Refs.~\cite{Bakoyeorgos1982,Paradellis1981}. The state at 1972 keV, which is strongly populated by the 746-keV transition, was found to decay via four $\gamma$ rays - an $E1$, high-intensity in-band 635-keV transition connecting to the 1337-keV level, an $E1$, 485-keV one feeding the state at 1488 keV, an $M2$, 1398-keV transition reaching the 574-keV level, and a newly observed 208-keV, $E1$ $\gamma$ ray connecting to the 1765-keV state. These placements, previously reported in Refs.~\cite{Bakoyeorgos1982,Paradellis1981}, are confirmed in the present work, and are demonstrated in Fig. \ref{fig:g636_747_916} (b) by a coincidence spectrum resulting from a double gate on the 746- and 916-keV $\gamma$ rays. Examining the spectra obtained from a double coincidence gate on the 635- and 746-keV $\gamma$ rays, presented in Fig. \ref{fig:g636_747_916} (a), reveals the presence of transitions such as the 602-, 671-, 694-, 824-, 1004-, and 1067-keV $\gamma$ rays populating the 2719-keV level. Notably, these transitions are absent in the energy spectrum gated on 746- and 916-keV $\gamma$ rays (see Fig. \ref{fig:g636_747_916} (b)). This absence supports their placement on top of the 2719-keV state, in parallel with the high-intensity 916-keV $\gamma$ ray. The 671-, 824-, and 1004-keV $\gamma$ rays were previously observed in Ref.~\cite{Bakoyeorgos1982}, and their placements are consistent with the present work. However, due to low statistics, their spins and parities were uncertain. In present work, the high statistics acquired with the Gammasphere detector array enabled the precise determination of their multipolarities. Therefore, the measured $R_{ac}$ values indicate a mixed $M1+E2$ character for the 671- and 824-keV $\gamma$ rays while the 1004-keV transition is of $E2$ nature. Furthermore, a mixed $M1+E2$ character is established for the newly identified 694-keV $\gamma$ ray, while the 1067-keV one connecting the ${15/2}^-$ level at 3786-keV to the ${13/2}^+$ state at 2719 keV, is assigned an $E1$ character. 

\begin{figure}[ht]
    \centering
    \includegraphics[width=\columnwidth]{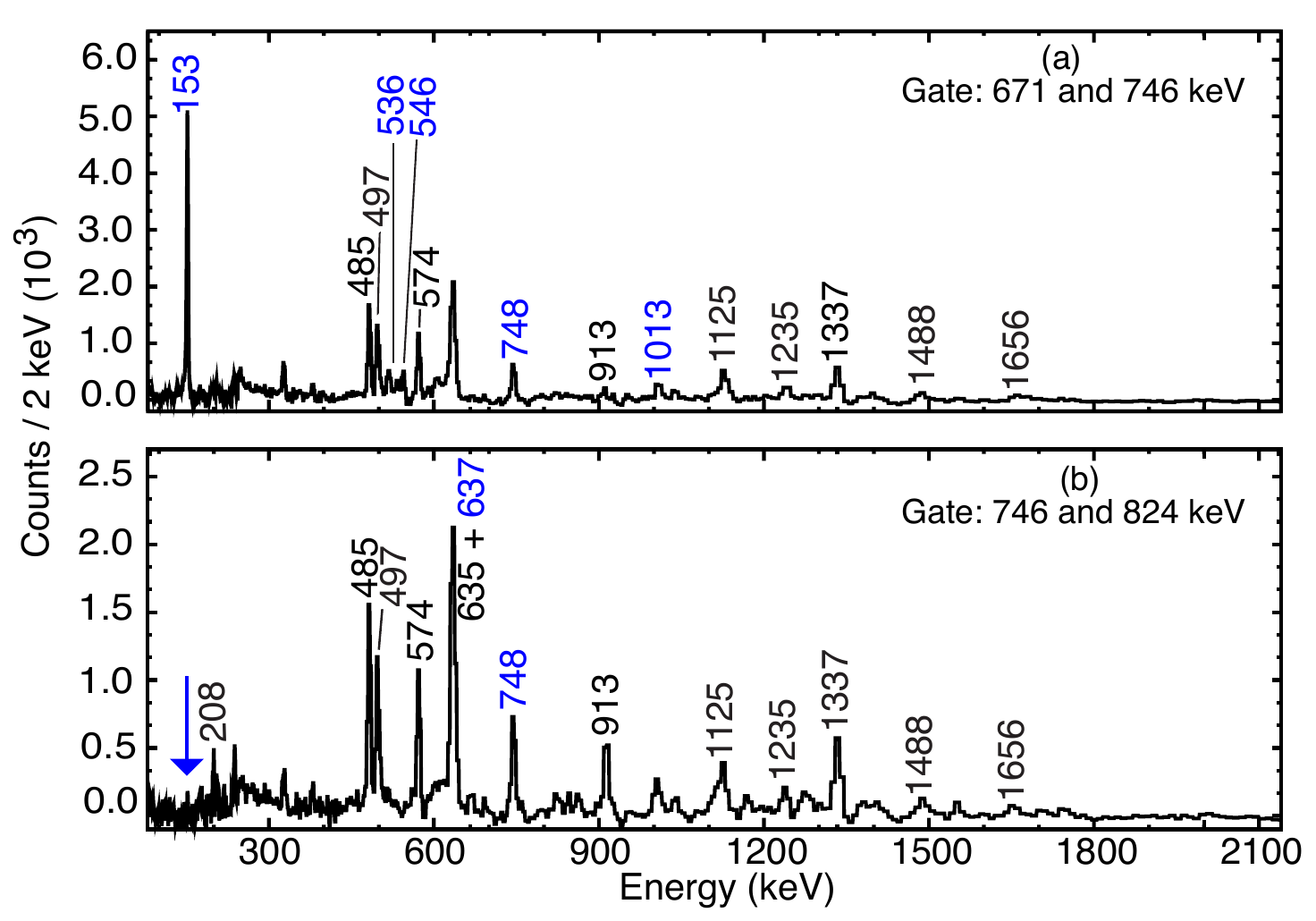}
    \caption{Coincidence spectra resulting from double coincidence gates on (a) 671- and 746-keV and (b) 746- and 824-keV $\gamma$ rays. The transitions decaying directly to the 3390-keV level are marked in blue color. The blue arrow in (b) marks the position of the 153-keV $\gamma$ rays that was not observed in coincidence with the 824-keV transition. (see text for details).}
    \label{fig:g671_747_824}
\end{figure}

The spectrum obtained from a double coincidence gate on the 746- and 671-keV $\gamma$ rays, presented in Fig.~\ref{fig:g671_747_824} (a), confirms the 546-536-1013-keV cascade connecting the ${23/2}^+$, 5485- with ${21/2}^+$, 4472-, and ${19/2}^+$, 3359-keV levels. Notably, the spectrum also reveals the presence of the 153-keV $\gamma$ ray as well as another 748-keV one in coincidence with the 746- and 671-keV gates. However, these two $\gamma$ rays are absent in the double-gated coincidence spectrum of the 746- and 916-keV transitions. Interestingly, the 746-keV line appears when gating on the 746- and 824-keV $\gamma$ rays, as seen in Fig.~\ref{fig:g671_747_824} (b). This confirms the presence of a 746-keV doublet, while the 153-keV $\gamma$ ray is notably absent, as indicated by the blue arrow. These observations further support the placement of the 153-keV transition between the ${15/2}^+$, 3543-, and ${15/2}^+$, 3390-keV levels, as well as the placement of the 748-keV doublet atop the 4180-keV state.

Coincidence spectra resulting from double gates on 916- and 1168-keV as well as 1168- and 1179-keV transitions are displayed in panels (a) and (b) of Fig. \ref{fig:g916_1168_1179}. Panel (a) isolates the 1179-keV $\gamma$ ray, which depopulates the ${25/2}^+$, 5983-keV state free from contamination by the neighboring 1168-keV transition. Conversely, panel (b) allows extraction of the 1168-keV peak without interference from the 1179-keV $\gamma$ ray. The measured $R_{ac}$ values confirm an $E2$ character for both the 1168- and 1179-keV transitions. Note that the ${25/2}^+$, 5983-keV level is also populated by the 1125- and the low-intensity 1253-keV transitions and de-excites via two additional 248- and 497-keV $\gamma$ rays. The 248-keV transition exhibits an $E1$ character, and populates the 5734-keV state in band NP I, while the 497-keV $\gamma$ ray displays a mixed $M1+E2$ multipolarity, based on the angular distribution analysis and is associated with the feeding of the 5485-keV level. 

\begin{figure}[ht]
    \centering
    \includegraphics[width=\columnwidth]{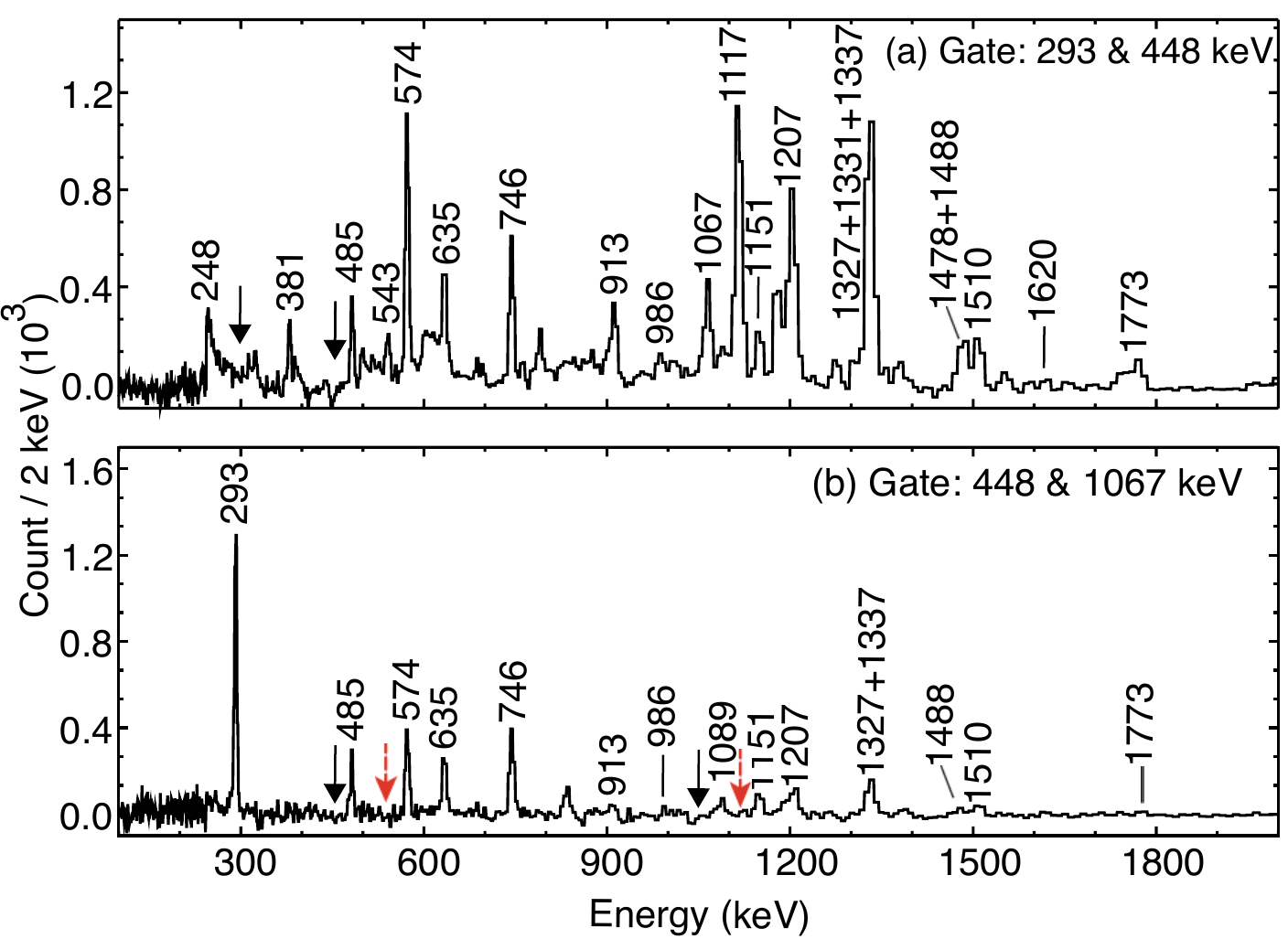}
    \caption{Coincidence spectra resulting from a double gate on the (a) 293- and 448-keV, and (b) 448- and 1067-keV $\gamma$ rays. The black arrows point to the transitions gated upon. The red ones point to the supposed to be missing 543- and 1117-keV transitions (See text for details).}
    \label{fig:g293_448_1067}
\end{figure}

\begin{figure}[ht]
    \centering
    \includegraphics[width=\columnwidth]{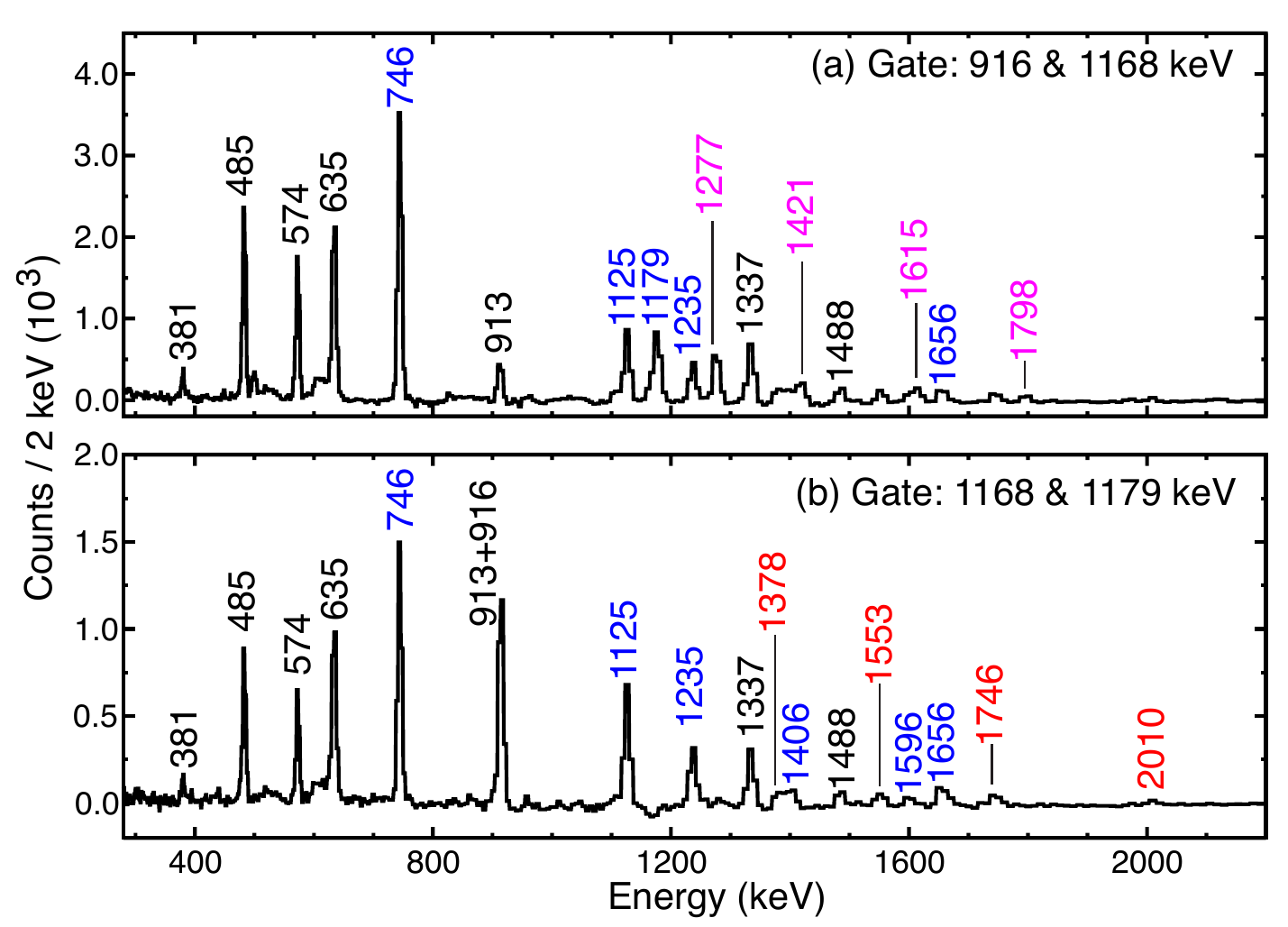}
    \caption{Coincidence spectra resulting from a double gate on the (a) 916-and 1168-keV, and (b) 1168- and 1179-keV transitions, showing transitions from the rotational-like sequences S I and S III in magenta and red, respectively.}
    \label{fig:g916_1168_1179}
\end{figure}

\begin{figure}[ht]
    \centering
    \includegraphics[width=\columnwidth]{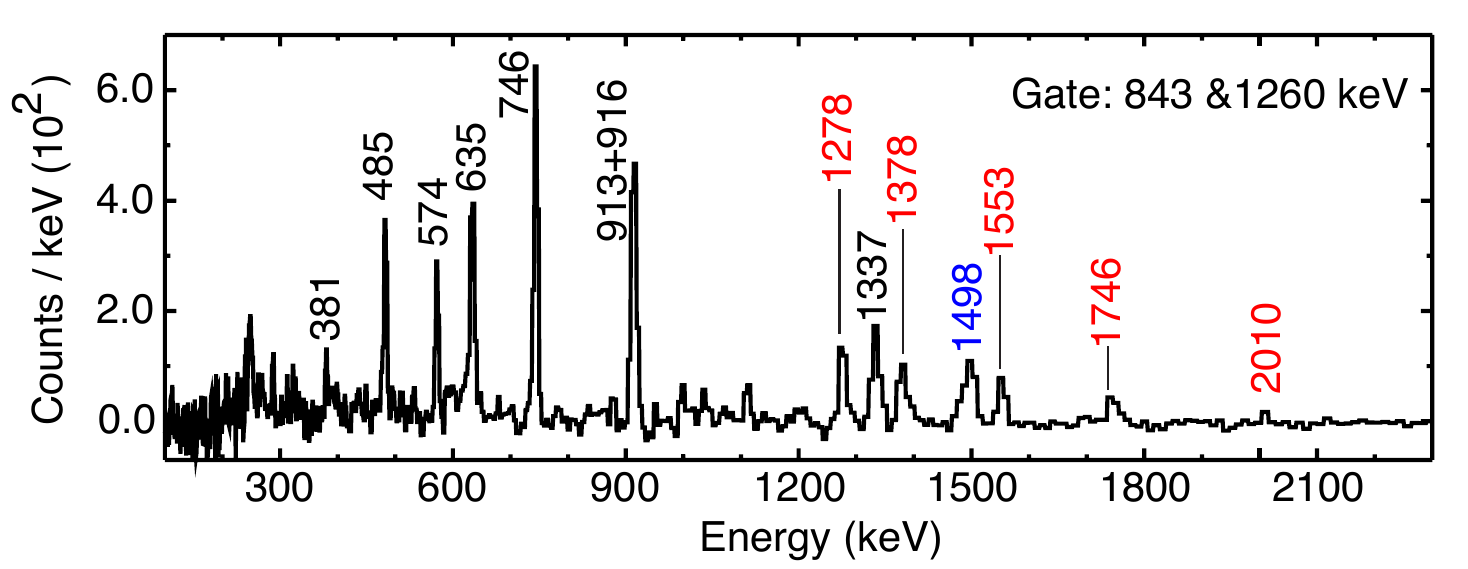}
    \caption{Coincidence spectra resulting from a double gate on the 843-and 1260-keV transitions, showing the onset of a rotational-like structure.}
    \label{fig:g843rb}
\end{figure}

\begin{figure}[ht]
    \centering
    \includegraphics[width=\columnwidth]{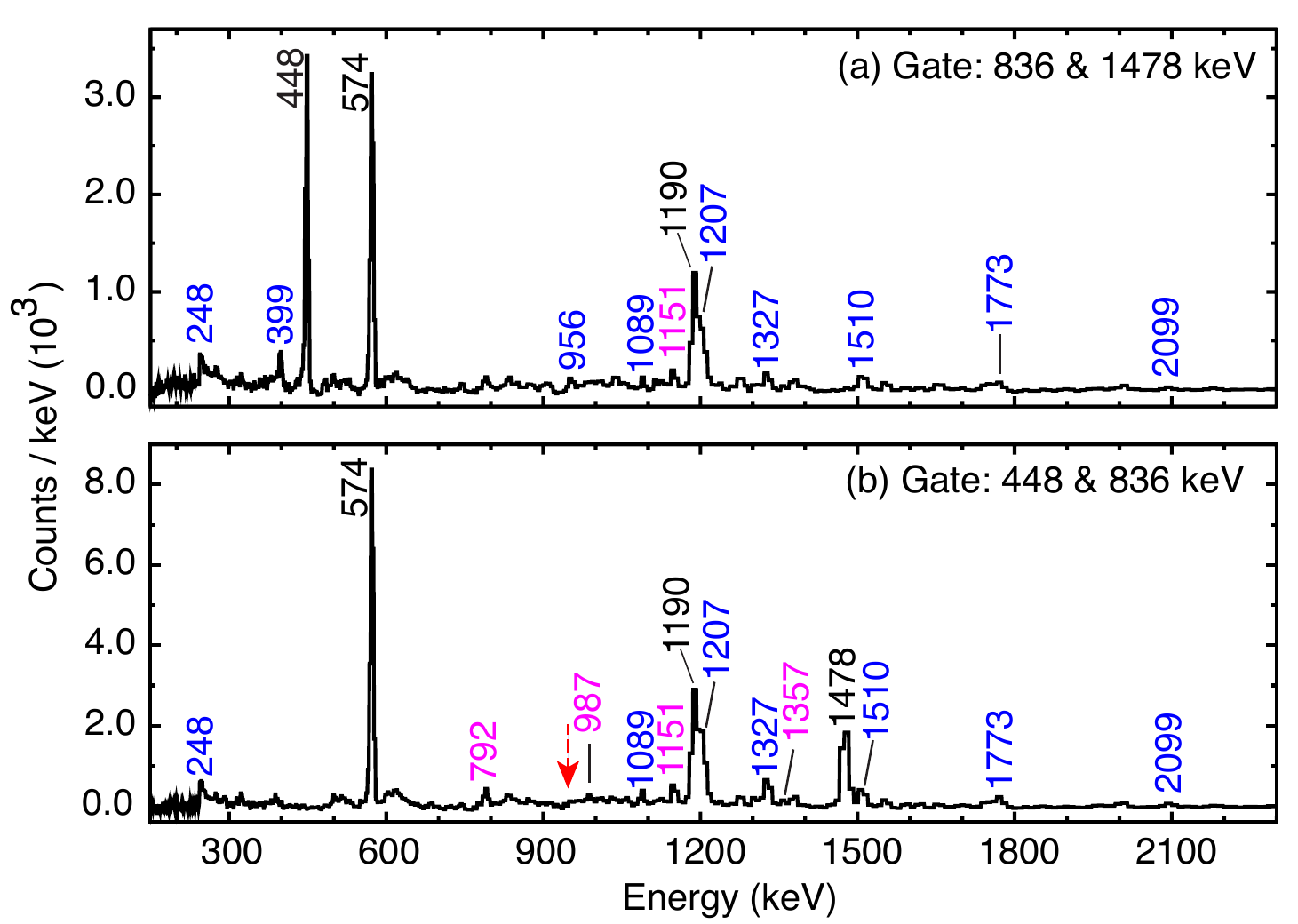}
    \caption{Gamma-ray spectrum resulting from a double coincidence gate on the (a) 836- and 1478-keV, and (b) 448- and 1478-keV $\gamma$ rays. Transitions within the side bands SBI, SBII and SBIII are marked in magenta. Other newly-placed transitions in the NPI sequence are in blue.} 
    \label{fig:g1478_836_448}
\end{figure}

A double coincidence gate on the 843- and 1260-keV transitions presented in Fig. \ref{fig:g843rb} (a) reveals the high-spin sequence (labeled as S III in Fig. \ref{fg:low_sch}) built atop the ${27/2}^+$, 7236-keV level, extending to the ${47/2}^+$, 15201-keV state via a 1278-1378-1553-1746-2010-keV cascade of $E2$ transitions, as confirmed by the measured $R_{ac}$ values. Another high-spin sequences (S II) is observed above the 4803-keV level, comprising 1277-1421-1615-1798-2115-(2224)-keV $E2$ cascade. Additionally, another cascade of six in-band $E2$ transitions forms sequence S I, which extends between the ${21/2}^+$, 4803- level and the $({45/2}^+)$, (14282)-keV state. The present work also identified an additional collective cascade (labeled NP I in Fig. \ref{fg:low_sch}) built on the 574-keV, first-excited state and extending it to $E_x = 10670$ keV with $I^\pi = {35/2}^-$. The negative parity of this sequence was initially proposed in \cite{Harms-Ringdahl1974} based on single-particle excitations in the $p_{3/2}$, $p_{1/2}$, and $f_{5/2}$ orbitals, and, later confirmed in \cite{Bakoyeorgos1982}. The $R_{ac}$ analysis of the newly-observed transitions within this sequence revealed mostly $E2$ character, enabling the determination of the spins and parities of the associated levels. With improved statistics, several side cascades, labeled as SBI, SBII, and SBIII in Fig. \ref{fg:low_sch} with transitions feeding into NP I have also been identified. Transitions within these newly-observed side feeding cascades are presented in the coincidence spectra of Figs. \ref{fig:g1478_836_448} (a) and (b). 

It should be noted that, while the present manuscript was in preparation, the authors became aware of the work reported in Ref.\cite{sharma2025aspectssingleparticleexcitations}, in which the structure of $^{69}$Ga was investigated using the $^{59}$Co($^{13}$C,2pn) reaction with the Indian National Gamma Array (INGA) of Compton-suppressed HPGe detectors. That study extended the known level scheme toward intermediate-spin states. The observed levels were interpreted in part through large basis shell-model calculations, while the band-like sequences were discussed in the framework of Total Routhian Surface (TRS) calculations, which highlighted the role of deformation in shaping the observed structures. Except for the multipolarity assignments of the 497-keV and 1007-keV transitions, which were identified in Ref.~\cite{sharma2025aspectssingleparticleexcitations} as having quadrupole and dipole character, respectively, in contrast to the dipole and quadrupole nature adopted in the present analysis, all the newly reported transitions in that work have been independently observed and placed in a similar manner in this study. The present results, obtained using a different reaction mechanism and complementary experimental conditions, not only confirm the majority of the placements reported in Ref.~\cite{sharma2025aspectssingleparticleexcitations} but also provide additional insights into the configuration assignments and decay pattern. In particular, the multipolarity analysis and coincidence relationships established here offer improved constraints on the spin assignments and support a more consistent interpretation of the observed band structures within the framework of quasi-particle excitations built on deformed configurations.

\section{DISCUSSION}\label{discussion}

To further investigate the properties of the level structure observed in $^{69}$Ga, shell-model calculations were carried out using the NUSHELLX code \cite{BROWN2014}. These calculations employed the JUN45 \cite{Honma2005, honma2009} and jj44b \cite{Lisetskiy2004} effective interactions within the $fpg$ shell-model valence space, which comprises the $0f_{5/2}$, $1p_{3/2}$, $1p_{1/2}$, and $0g_{9/2}$ proton and neutron orbitals. With an inert $^{56}$Ni core, this model space allows for excitations of three protons and ten neutrons within the valence shell, with the exclusion of core excitations.
Both these interactions have been successfully applied to describe level structures in several nuclei in this mass region including $^{71-78}$Ga \cite{Srivastava2012,Vedia2017,Zhong2021}, $^{74,76}$Ge \cite{Johnson2023,Ayangeakaa2023apr, Mukhopadhyay2017}, $^{62-64,70}$Ni \cite{Albers2016,albers63Ni,littleNi64,chiara2015} and $^{66,73}$Zn \cite{Ayangeakaa2022,Vedia2017}.

\begin{figure}[ht]
	\centering
    \includegraphics[width=\columnwidth]{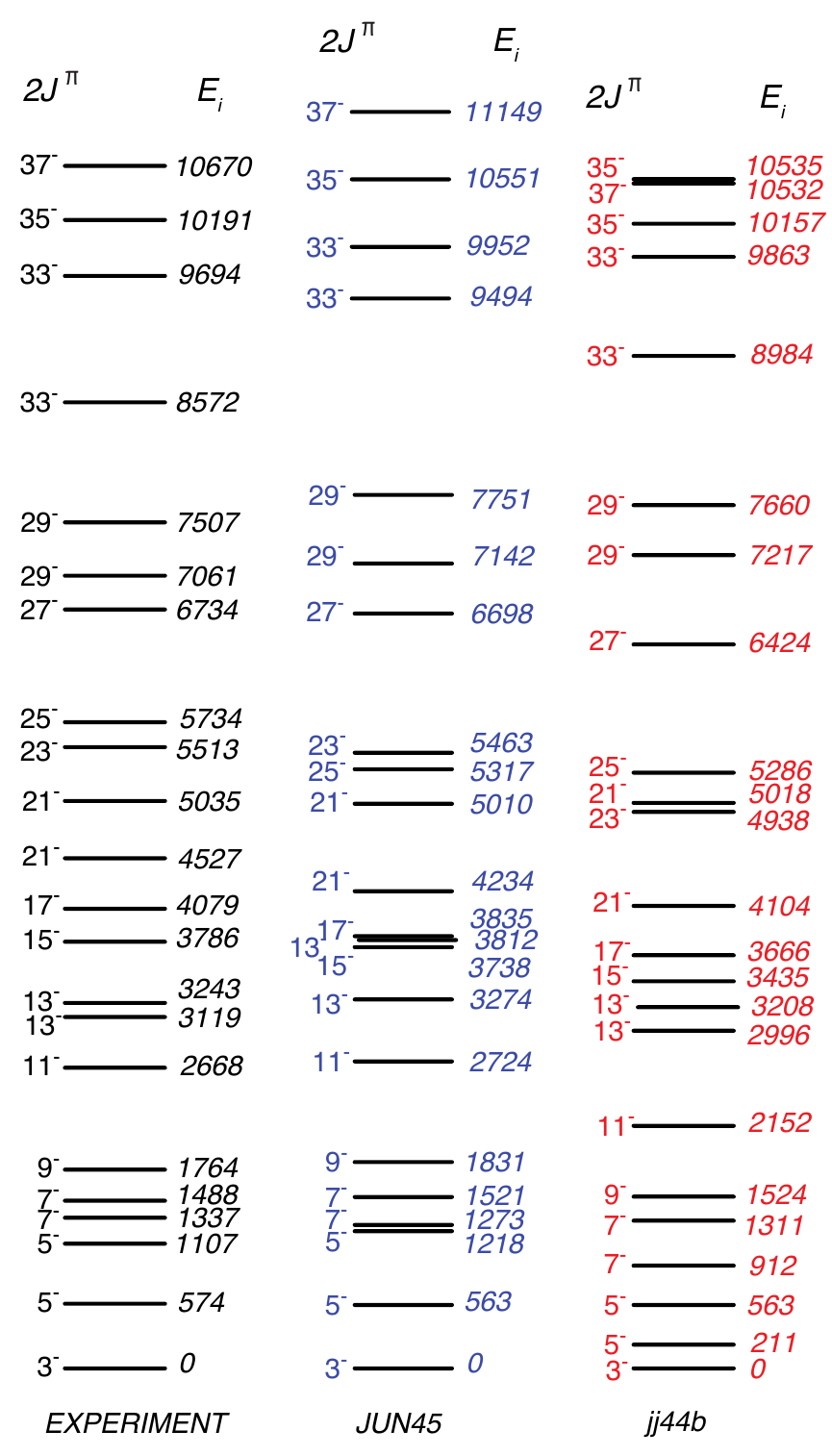}
	\caption{Comparison of experimental energy levels (in black) with the calculated ones using the JUN45 (in blue) and jj44b (in red) shell-model interactions for the negative parity-states in $^{69}$Ga}.
	\label{fg:Gajj44b_jun45-v1}
\end{figure}

\begin{figure}[ht]
	\centering
    \includegraphics[width=\columnwidth]{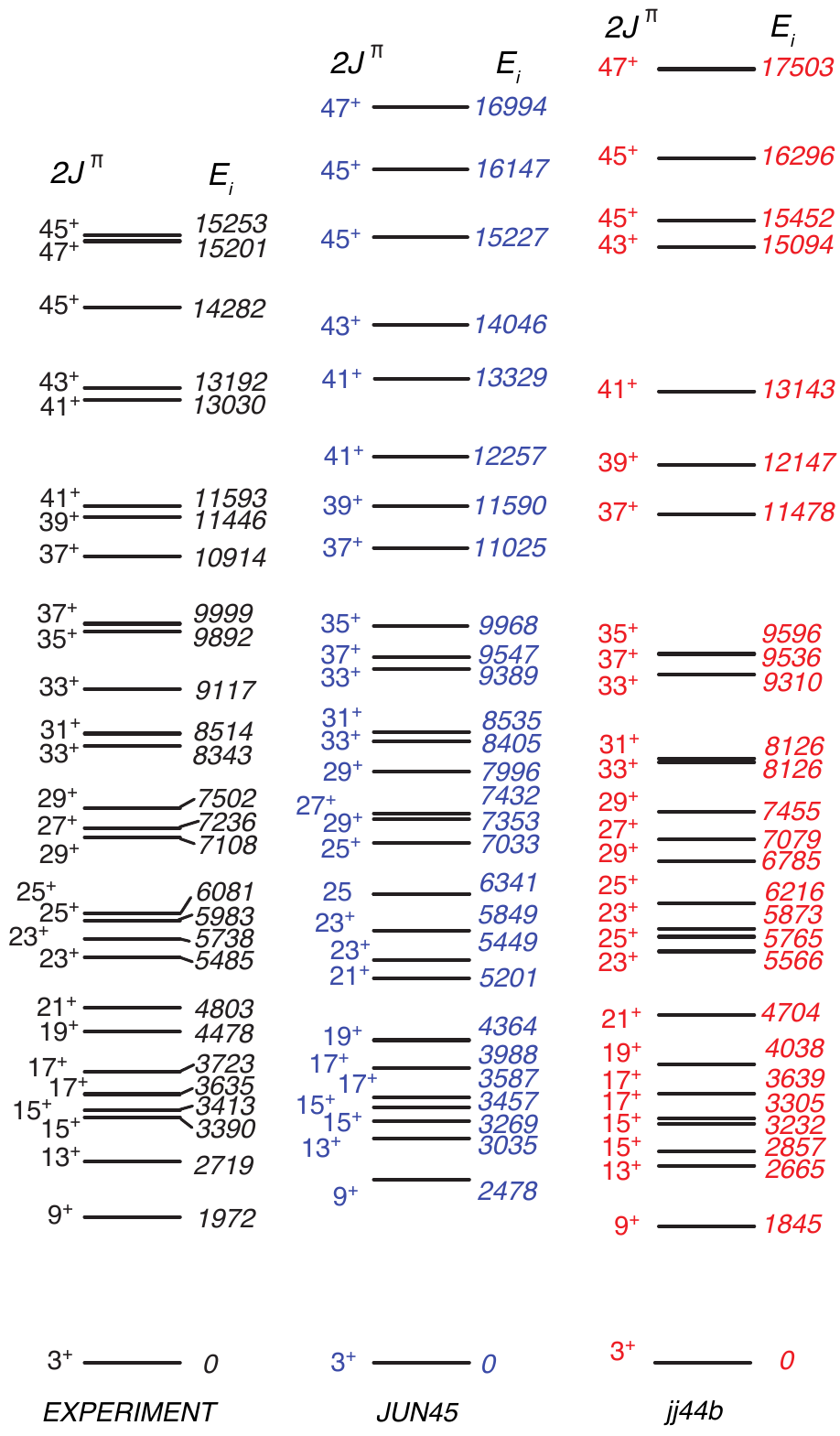}
	\caption{Comparison of experimental energy levels (in black) with the calculated ones using the JUN45 (in blue) and jj44b (in red) shell-model interactions for the positive-parity states in $^{69}$Ga}. 
	\label{fg:Gajj44b+jun45-v1}
\end{figure}

\begin{figure}[ht]
	\centering
    \includegraphics[width=\columnwidth]{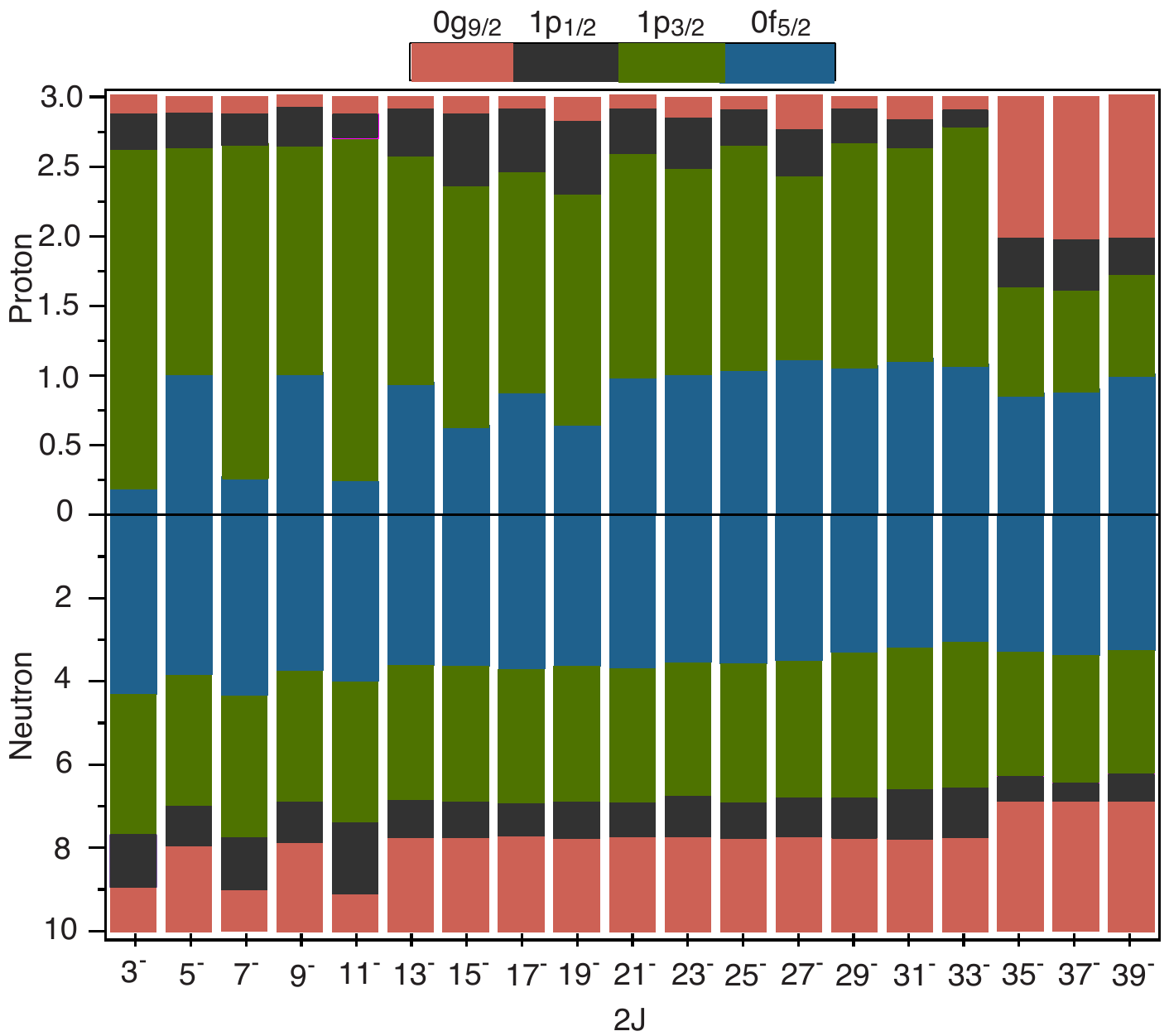}
	\caption{Proton and neutron occupation probabilities of the $f_{5/2}$, $p_{3/2}$, $p_{1/2}$, and $g_{9/2}$ orbitals for the negative-parity states in $^{69}$Ga using the JUN45 interaction.}
	\label{fg:jun45-occ}
\end{figure}

\begin{figure}[ht]
	\centering
    \includegraphics[width=\columnwidth]{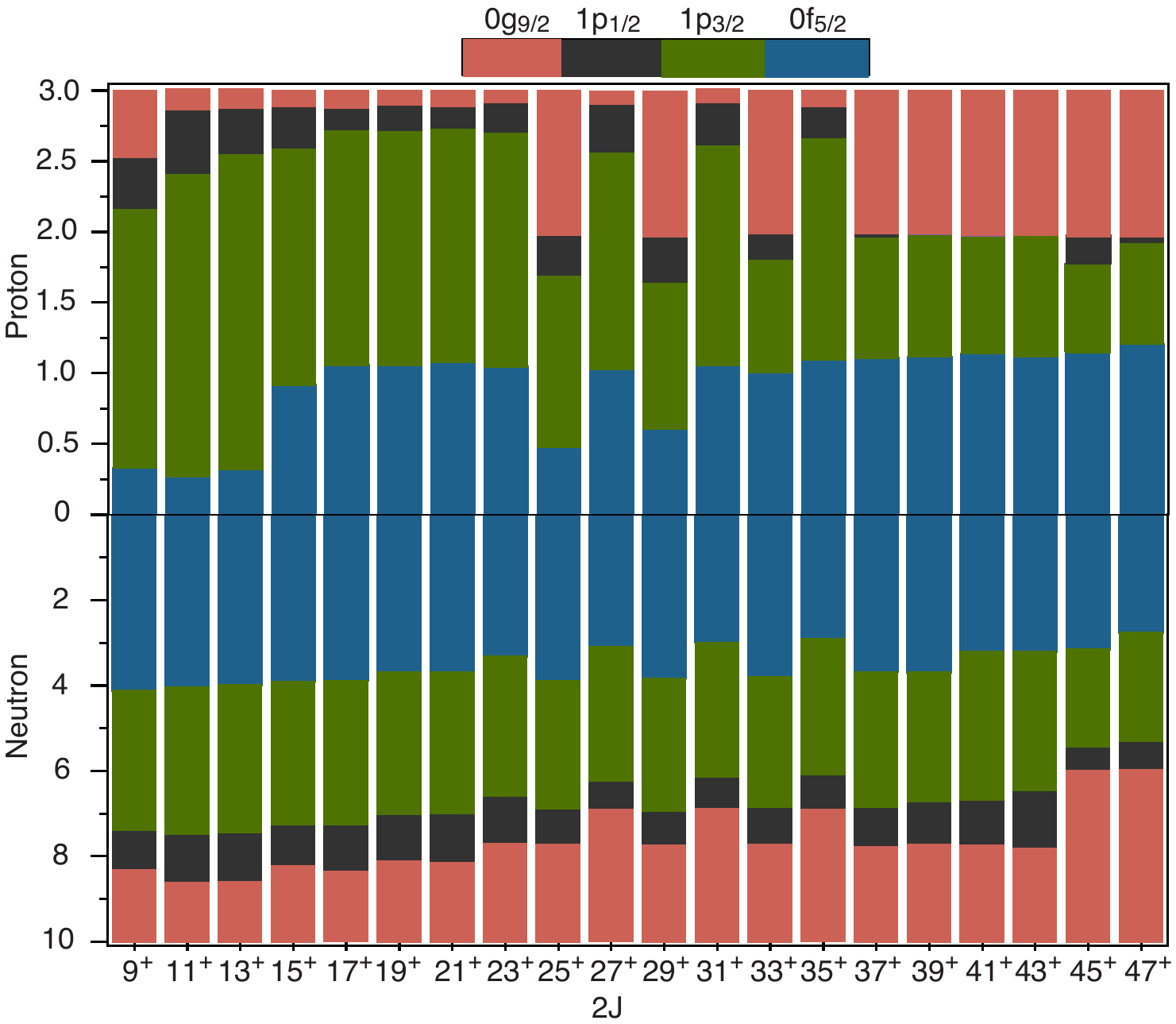}
	\caption{Proton and neutron occupation probabilities of the $f_{5/2}$, $p_{3/2}$, $p_{1/2}$, and $g_{9/2}$ orbitals for the positive-parity states in $^{69}$Ga using the JUN45 interaction.}
	\label{fg:jun45+occ}
\end{figure}

The results of the shell-model calculations, compared with the experimental data obtained in this study, are presented in Figs. \ref{fg:Gajj44b_jun45-v1} and \ref{fg:Gajj44b+jun45-v1}. The negative-parity states identified in the $^{69}$Ga nucleus primarily arise from proton quasiparticles in the $p_{3/2}$ orbital and unpaired neutrons in the $f_{5/2}$ ones, while the positive-parity states are dominated by contributions from the $g_{9/2}$ orbital. Proton and neutron occupation numbers for these states were also calculated using the shell model, and are presented in Figs. \ref{fg:jun45-occ} and \ref{fg:jun45+occ}. Both interactions accurately predict the ${3/2}^-$ ground-state spin and parity. Further analysis of the proton-neutron occupancy reveals a dominant $\pi (p_{3/2})^2 \otimes v[(f_{5/2})^4 (p_{3/2})^3 (p_{1/2})^1 (g_{9/2})^1]$ configuration for the ${3/2}^-$ ground state, with a proton occupation probability of 81\% in the $p_{3/2}$ orbital. The JUN45 interaction is found to be in better agreement with experimental data for the negative-parity states than the jj44b one. For example, the excited ${5/2}^-_1$ state is predicted by JUN45 to be only 11 keV below the measured value, with a dominant $\pi [(f_{5/2})^1 (p_{3/2})^2] \otimes v[(f_{5/2})^4 (p_{3/2})^3 (g_{9/2})^2]$ configuration, whereas jj44b underestimates the position of this level by 363 keV. Similarly, the second excited ${5/2}^-_2$ state at 1107 keV was computed by JUN45 to be 111 keV below the experimental value, with a dominant $\pi (p_{3/2})^2 \otimes v[(f_{5/2})^4 (p_{3/2})^3 (p_{1/2})^1 (g_{9/2})^1]$ configuration. In contrast, the jj44b Hamiltonian significantly underestimates this state, placing it 544 keV below the measured value.

In order to further quantify the disparities between data and computed excitation energies, the root-mean-square, $\sigma_{rms}$, deviation was calculated with the expression \cite{Xu1989,Steppenbeck2012}, 
$$\sigma_{rms}=\sqrt{\sum_{i=1}^N(E_{Cal}-E_{Exp})_i^2/N}$$\\ where $E_{Cal}$ and $E_{Exp}$ are, respectively, the calculated and the experimental energies for the $i^{th}$ level while $N$ represents the total number of states included in the calculations. For the first ten experimental levels, ranging from $I^\pi$ = ${5/2}^-$ to ${13/2}^-$ with excitation energies ranging from $\sim$ 0.5 to 3.0 MeV (Fig. \ref{fg:Gajj44b_jun45-v1}), the $\sigma_{rms}$ deviations from the JUN45 and jj44b interactions are $\sim$ 90 and 140 keV, respectively. However, upon including all experimental excited states below $\sim$ 6.0 MeV (with the exception of the ${3/2}^-$ ground state), the $\sigma_{rms}$ deviation increases to $\sim$ 340 and $\sim$ 600 keV for the JUN45 and jj44b interactions, respectively. This substantial increase in $\sigma_{rms}$ for the jj44b interaction can be attributed to the fact that its effective Hamiltonian was derived by specifically excluding the $^{69}$Ga experimental data from the iterative fit \cite{Lisetskiy2004}. 

The four lowest ${7/2}^-$ levels are better reproduced by JUN45 with deviations of 64, 33, 64, and 44 keV, respectively, while jj44b significantly underestimated these states by $\sim$ 425–665 keV. The configurations predicted by the JUN45 Hamiltonaian for these levels primarily involve proton and neutron excitations in the $p_{3/2}$, $f_{5/2}$, and $g_{9/2}$ orbitals (Fig. \ref{fg:jun45-occ}). The ${9/2}^-$ state at 1764 keV is well reproduced by JUN45, but is underestimated by jj44b by 240 keV, with both interactions, however, agreeing on a $\pi [(f_{5/2})^1(p_{3/2})^2]\otimes v[(f_{5/2})^4(p_{3/2})^3(g_{9/2})^2]$ configuration. Similarly, both interactions predict the two lowest ${13/2}^-$ states to decay via $E2$ transitions to the ${9/2}^-$ level, which further de-excites to the ${5/2}^-_1$ state by an $E2$ transition, consistent with the behavior of the yrast band. The ${11/2}^-_1$ state at 2574 keV is also better reproduced by JUN45, while the ${11/2}^-_2$ level is accounted for by both interactions. Intermediate states at $I^\pi$ = ${15/2}^-$ to ${29/2}^-$ are observed to be predominantly excitations involving the $\pi p_{3/2}$ and $v(f_{5/2}, p_{3/2}, g_{9/2})$ orbitals. The $\sigma_{rms}$ deviations for this group of states are $\sim$ 680 keV for JUN45 Hamiltonian and $\sim$ 840 keV for the jj44b one. Note that $\sigma_{rms}$ deviations of the order of $\sim$ 200 keV or less are generally considered to correspond to satisfactory agreement between the data and calculations. For positive-parity states, both the JUN45 and jj44b interactions reproduce the experimental levels well up to $I^{\pi}={25/2}^+$ with $\sigma_{rms}$ of the order $\sim$ 260 keV. The negative-parity states can be understood within the framework of configuration interaction shell model as well. The ${9/2}^+$ bandhead and the low-lying positive-parity states are calculated to be dominated by $\pi (p_{3/2}, g_{9/2})^3 \otimes v[(f_{5/2}, p_{3/2}, g_{9/2})]$ configurations. However, above the ${25/2}^+$ state, deviations increase significantly (towards $\sigma_{rms} \sim 1.0$ MeV), suggesting the onset of new modes of excitation due to significant proton and neutron occupation of the intruding $g_{9/2}$ orbital, as indicated in Figs. \ref{fg:jun45-occ}–\ref{fg:jun45+occ}. This suggests a structural change with increasing angular momentum, likely arising from enhanced collectivity.



\begin{figure}[ht]
	\centering
    \includegraphics[width=\columnwidth]{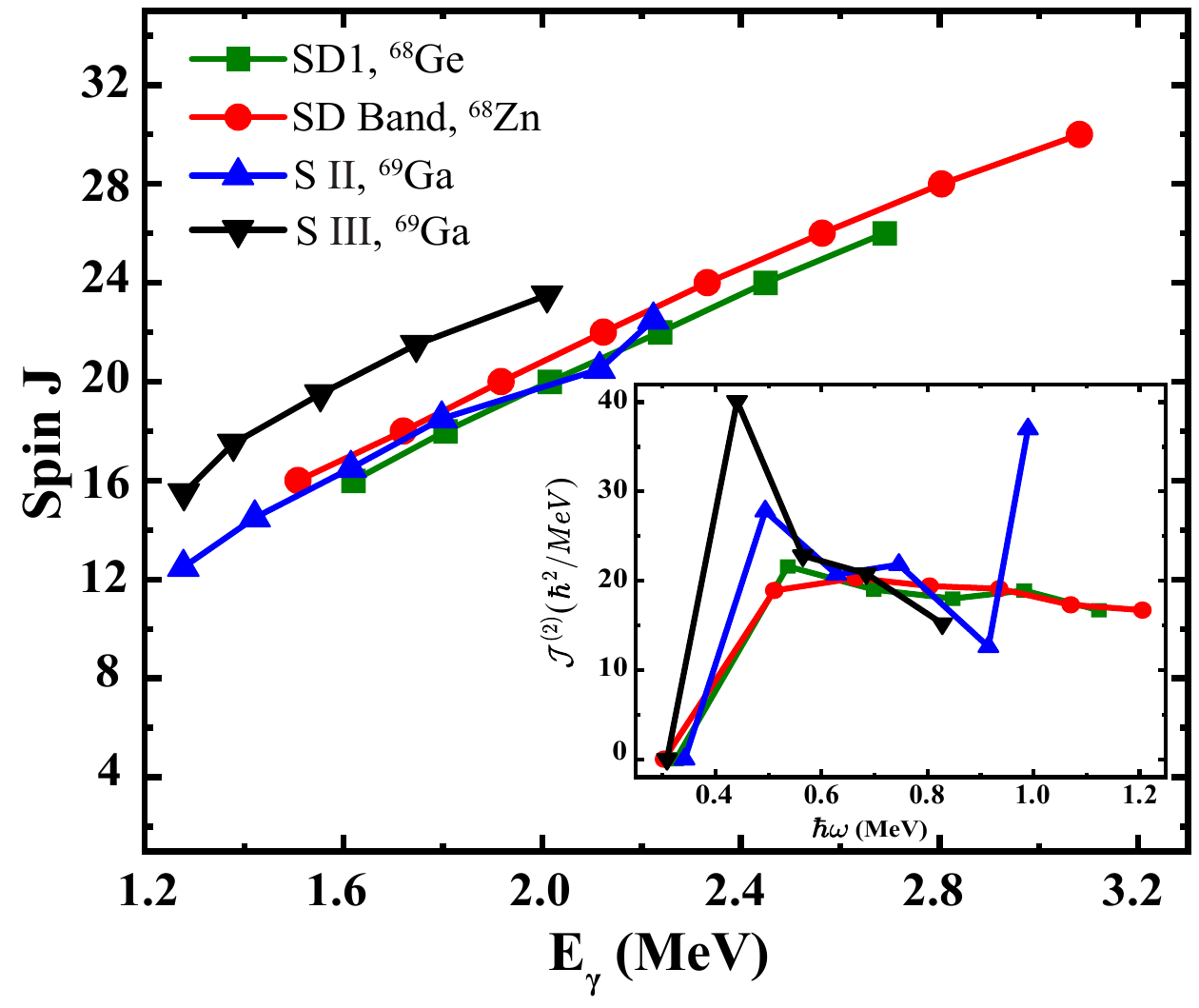}
	\caption{Angular momentum as a function of $\gamma-$ray energy with the insert in the lower-right panel showing the dynamic moments of inertia as a function of rotational frequency for the rotational bands in $^{68}$Ge and $^{68}$Zn, compared to the newly identified rotational-like sequences in $^{69}$Ga, S II and S III.}
	\label{fg:SPINvsEg-v1}
\end{figure}

Figure \ref{fg:SPINvsEg-v1} compares the evolution of angular momentum versus \(\gamma\)-ray energy for the newly-identified rotational-like sequences (S II and S III in Fig.~\ref{fg:low_sch}), with that seen in well-established rotational bands observed in neighboring \(^{68}\)Ge \cite{68ge} and in the \(^{68}\)Zn  isotone \cite{68zn}. For collective rotational sequences, a constant slope corresponds to a constant dynamic moment of inertia, \(J^{(2)}\) \cite{BENGTSSON198514}. However, for S II and S III in $^{69}$Ga, the plot is nearly curved, pointing to a variation in the moment of inertia, herewith suggesting intrinsic configurations or interactions in $^{69}$Ga that are likely different from those in neighboring $^{68}$Ge and $^{68}$Zn. In addition, the dynamic moment of inertia, $J^{(2)}$ has been plotted as a function of rotational frequency for the sequence S II and S III of \(^{69}\)Ga compared to the rotational bands in \(^{68}\)Ge and \(^{68}\)Zn in the lower right insert in Fig. \ref{fg:SPINvsEg-v1}. Superdeformed bands in $^{68}$Ge and $^{68}$Zn demonstrate a nearly constant moment of inertia, and are understood to originate from broken $^{56}$Ni core with excitation of $2p-2h$ from the $\pi f_{7/2}$ orbital into the $g_{9/2}$ intruder orbital \cite{68ge}. However, the rotational-like sequences in $^{69}$Ga exhibit a decreasing trend in $J^{(2)}$, with a sharp increase at $hw\sim0.4$ MeV, probably due to the alignment of protons and neutrons or nucleon interactions \cite{65zn}. Note that the sharp increase of \(J^{(2)}\) in S II at $hw\sim1.0$ MeV cannot be certainly characterized as alignment due to insufficient data at this high spin.

To further investigate the collective behavior of the newly observed structures in $^{69}$Ga, theoretical calculations based on the tilted-axis cranking covariant density functional theory (TAC-CDFT) framework have been utilized, as reported in Ref.\cite{wang2025tac-cdft}. These calculations \footnote{Selected results from Ref.~\cite{wang2025tac-cdft} are reproduced in this manuscript with permission from the corresponding authors.} were performed independently and provide theoretical predictions for rotational sequences built on deformed configurations involving the $g_{9/2}$ orbital. In the calculations, the shell-model-like approach (SLAP)~\cite{ZengJY1983NPA,MengJ2006FPC,WangYP2024PRL,XuFF2024PRL} was employed to treat pairing correlations while conserving particle number, an essential feature for accurately describing rotational phenomena in nuclei. The calculations adopted a monopole pairing interaction in the particle-particle channel, with effective neutron and proton pairing strengths of 0.7 and 0.75 MeV, respectively, determined by reproducing empirical odd-even mass differences. The PC-PK1 relativistic density functional~\cite{ZhaoPW2010PRC} was used in the particle-hole channel; this functional has been validated across a wide range of structure phenomena including nuclear masses~\cite{YangYL2021PRC,ZhangKY2022ADNDT}, nuclear shapes~\cite{XuFF2024PRC,XuFF2024PLB,YangYL2023PRC}, and various types of rotational manifestations  such as magnetic and antimagnetic~\cite{ZhaoPW2011PRL,ZhaoPW2011PLB,MengJ2016PS,WangYK2017PRC,WangYK2018PRC}, chiral~\cite{ZhaoPW2017PLB,WangYK2019PRC,RenZX2020PRC,WangYP2023PLB}, and fission dynamics~\cite{RengZX2022PRL,LiB2023PRC}. The Dirac equation for single-particle states was solved using a three-dimensional harmonic oscillator basis comprising ten major shells, providing convergence for nuclei in the $A \sim 60$ region~\cite{ZhaoPW2011PLB}. The many-particle configuration (MPC) space was truncated at an excitation energy cutoff of $E_c = 10$ MeV~\cite{WuCS1989PRC}, and convergence of the results was confirmed by testing with a larger configuration space and renormalized pairing strength.

\begin{figure}[htbp!]
  \centering
  \includegraphics[width=\columnwidth]{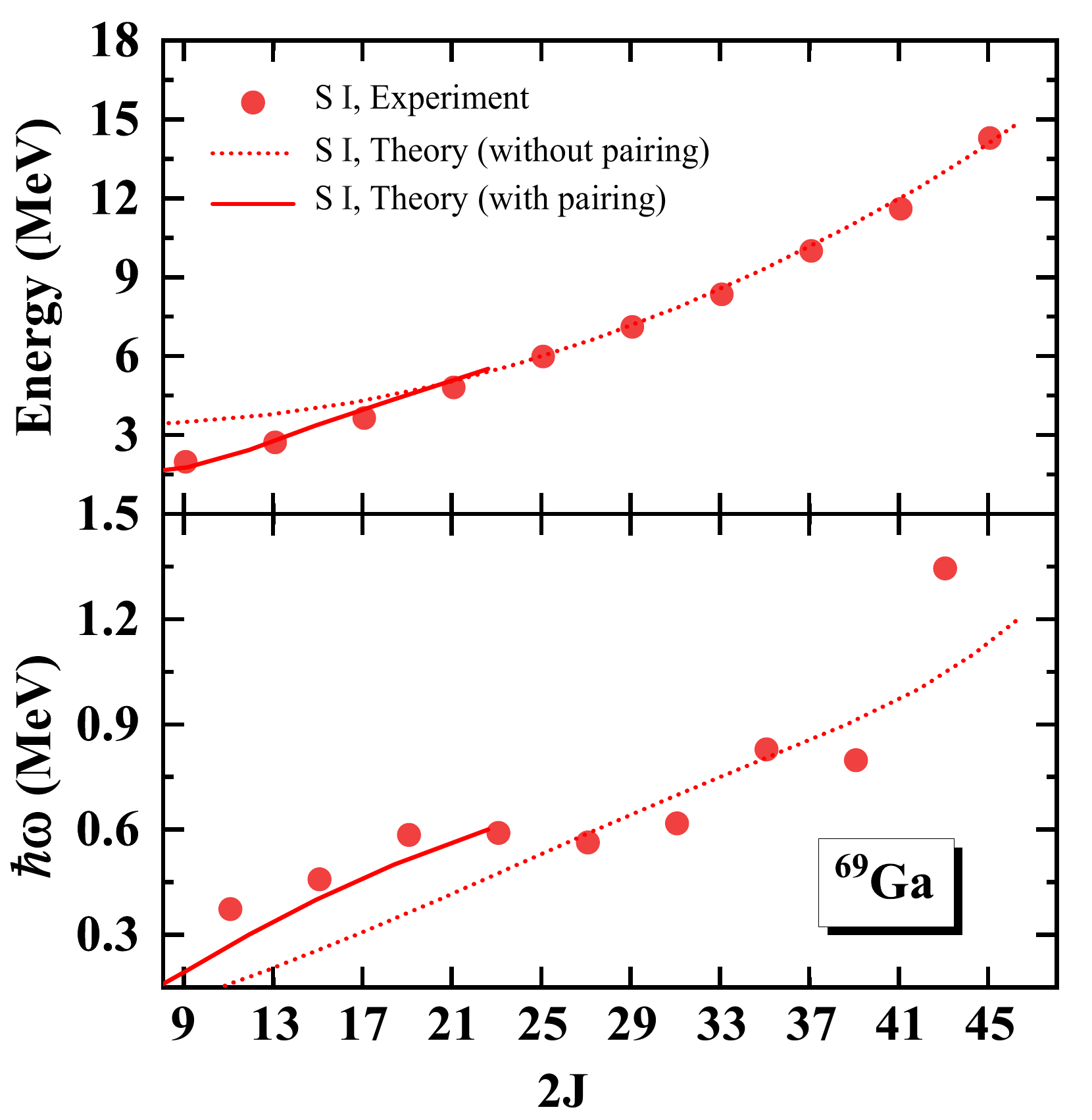}
  \caption{Energy (upper panel) and rotational frequency (lower panel) as functions of the angular momentum for the positive-parity sequence S I in $^{69}$Ga calculated by TAC-CDFT-SLAP with and without pairing, in comparison with the data. See text for details. Reproduced from Ref.~\cite{wang2025tac-cdft} with permission.}
  \label{fig1}
\end{figure}

\begin{figure}[htbp!]
  \centering
  \includegraphics[width=\columnwidth]{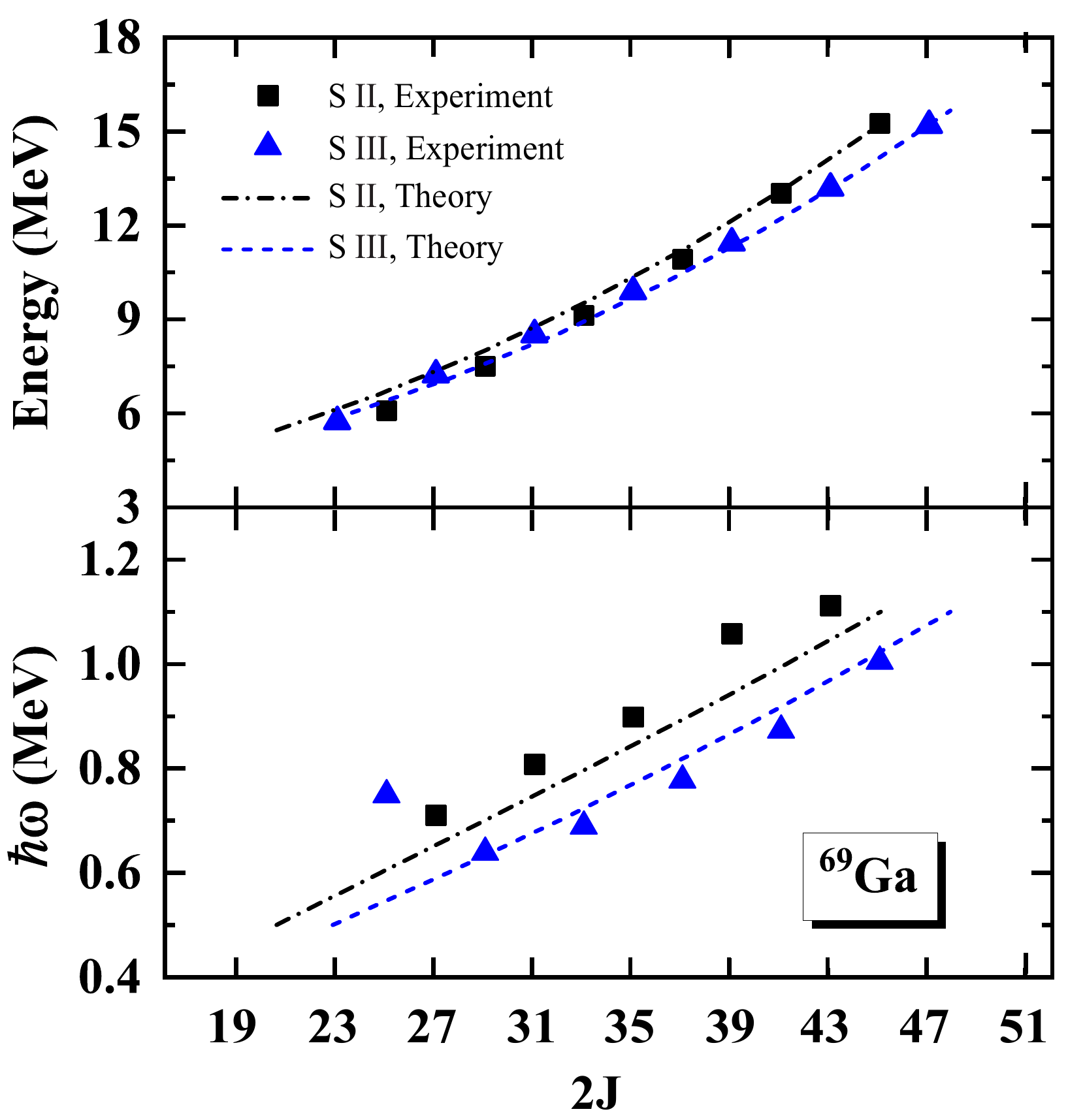}
  \caption{Energy (upper panel) and rotational frequency (lower panel) as functions of the angular momentum for the positive-parity sequences S II and S III in $^{69}$Ga calculated by TAC-CDFT-SLAP without pairing, in comparison with the data. Reproduced from Ref.~\cite{wang2025tac-cdft} with permission.}
  \label{fig2}
\end{figure}

The occupation of the $g_{9/2}$ orbitals plays a pivotal role in describing the collective structures observed in $^{69}$Ga. To identify suitable positive-parity configurations involving these $g_{9/2}$ orbitals, cranking calculations using the configuration-fixed constrained approach \cite{MengJ2006PRC} were performed. By comparing the calculated results with experimental data, the valence-nucleon configuration for sequence S I was determined to be $\nu(1g_{9/2})^4(2p_{3/2},1f_{5/2})^6\otimes \pi(1f_{7/2})^{-2}(1g_{9/2})^1$. Similarly, the configuration $\nu(1g_{9/2})^4(2p_{3/2},1f_{5/2})^6\otimes\pi(1g_{9/2})^3$ with positive and negative signatures were assigned to S II and S III, respectively. These configurations generate principal axis rotations, consistent with the observed $\Delta$I = 2 character of the bands.

The calculated energy spectrum and angular momenta for the positive-parity band S I are presented in Fig. \ref{fig1}, along with the experimental data. The theoretical results reproduce the experimental values well, particularly for states with spin $I \leq 21/2$, where pairing correlations are crucial. Without accounting for pairing, the calculations tend to overestimate both the energy spectrum and the angular momenta for a given rotational frequency. Analyzing angular momentum alignments reveals that the unpaired-nucleon configuration for $I \leq 21/2$ is $\pi(g_{9/2})^1$. At higher spins, two neutrons in the $1g_{9/2}$ orbitals align, resulting in the unpaired-nucleon configuration $\nu(1g_{9/2})^2\otimes\pi(1g_{9/2})^1$. This alignment suggests a possible backbending phenomenon around $I = 25/2$. Due to the significant reduction in neutron pairing correlations caused by the blocking of $1g_{9/2}$ orbitals, calculations for higher-spin states were performed without pairing, and the agreement with experimental data is satisfactory (see Fig. \ref{fig1}).

Figure \ref{fig2} compares the calculated energy spectrum and angular momenta for the positive-parity sequences S II and S III with experimental data. The TAC-CDFT-SLAP calculations are in close agreement with the data, supporting the interpretation of these sequences as possible signature partners. By examining angular momentum alignments, the unpaired-nucleon configuration in the low-spin region is identified as $\nu(1g_{9/2})^2\otimes \pi(1g_{9/2})^1$, with the unpaired proton occupying the $1g_{9/2}$ orbital and having an approximate value of $\Omega\approx3/2$ for the projection of the total angular momentum along the shape's symmetry axis. This differs from S I, where the unpaired proton occupies the $1g_{9/2}$ orbital with $\Omega\approx1/2$. At higher spins, two additional valence $g_{9/2}$ protons begin to contribute significantly to the angular momentum, yielding the unpaired-nucleon configuration $\nu (1g_{9/2})^2\otimes \pi(1g_{9/2})^3$. Additional experimental work will be required to substantiate this interpretation further. For example, an extension of the three sequences toward higher spins is desirable as well as information on state lifetimes, which would provide new insight into the degree of collectivity.

\section{SUMMARY AND CONCLUSIONS}\label{summary}
Intermediate and high-spin states in odd-$A$ $^{69}$Ga were investigated using the $^{26}$Mg($^{48}$Ca, $p4n$) reaction. The measurements were performed at the ATLAS accelerator facility using the Gammasphere array coupled to the FMA. The low-spin structure of $^{69}$Ga, presented in \cite{Bakoyeorgos1982}, consisting of both negative- and positive-parity states, has been significantly expanded from $\sim$ 4.5 to $\sim$ 15 MeV, based on coincidence relationships. Angular distribution and correlation measurements were performed to determine the multipolarity of the newly observed $\gamma$ transitions. Three high-spin sequences with quadrupole transitions in coincidence with the low-spin states were observed. Shell-model calculations performed with the JUN45 and jj44b effective interactions in the $f_{5/2}pg_{9/2}$ valence space have been successfully employed to describe the low-spin states. For the states at high spin, calculations based on the tilted-axis-cranking covariant density functional theory (TAC-CDFT) were performed. For the yrast sequence S I, pairing correlations play a crucial role in the description of states with spin \( I \leq 21/2\). Also, the alignment of two $1g_{9/2}$ neutrons suggests a possible backbending effect around $I = 25/2$. The calculated energy and angular momenta for positive-parity sequences S II and S III in $^{69}\text{Ga}$ are closely reproduced by TAC-CDFT-SLAP calculations. These bands are suggested to be signature partners. In the lowest parts of these sequences, the unpaired nucleon configuration is \( \nu(1g_{9/2})^2 \otimes \pi(1g_{9/2})^1 \), with the unpaired proton occupying the \( 1g_{9/2} \) orbital with \( \Omega \approx 3/2 \), which differs from that of S I, where the unpaired proton occupies the same orbital with \( \Omega \approx 1/2 \). In conclusion, the observation of possible signature partner, and angular momentum alignment phenomena resulting from two neutrons in the $g_{9/2}$ orbital are important as these provide insight into the evolution of single-particle excitations as a function of angular momentum, and the emergence of collective behavior at high spins near \( N = 40 \), arising from the influence of the shape-driving \( \pi g_{9/2} \) and \( \nu g_{9/2} \) intruder orbitals.

\section{Acknowledgements}
The authors thank Y. P. Wang, Y. K. Wang, and P. W. Zhao, the contributors of Ref.~\cite{wang2025tac-cdft}, for granting permission to reproduce selected results from their calculations in this manuscript. This material is based upon work supported by the U.S. Department of Energy, Office of Science, Office of Nuclear Physics, under Grants No. DE-SC0023010 (UNC), DE-FG02-97ER41041 (UNC), No. DE-FG02-97ER41033 (TUNL), No. DE-FG02-94ER40848 (UML), Contracts No. DE-AC02-06CH11357 (ANL), No. DE-AC02-98CH10886 (BNL), Grant No. DEFG02-94-ER40834, NSF Grant No. PHY-2208137. This research uses resources of ANL's ATLAS facility, which is a U. S. Department of Energy Office of Science User facility. The views expressed in this article are those of the authors and do not reflect the official policy or position of the U. S. Naval Academy, Department of the Navy, the Department of Defense, or the U. S. Government.

\bibliography{references}

\end{document}